\documentclass[aps,pra,twocolumn,notitlepage,10pt,superscriptaddress,longbibliography,nofootinbib]{revtex4-2}

\usepackage{graphicx}
\usepackage{times,bbm,amsmath,amssymb}

\usepackage{hyperref}
\usepackage{xcolor}
\usepackage{cleveref}
\crefname{figure}{Fig.}{Figs.}
\Crefname{figure}{Figure}{Figures}

\usepackage{graphicx}
\usepackage{bm}
\usepackage{soul}
\usepackage{pifont}
\usepackage{bbold}
\usepackage{physics}

\usepackage{parskip}
\usepackage{libertine}[newtxmath]

\newcommand{\calO}{\mathcal{O}}

\newcommand{\bs}[1]{\boldsymbol{#1}}

\newcommand{\bscalO}{{\bs\calO}}
\newcommand{\ntr}{{n_{\rm tr}}}

\newcommand{\nobs}{{n_{\rm obs}}}

\definecolor{azure}{rgb}{0.0, 0.3, 1.0}

\begin{document}

\title{Efficient classical training of model-free quantum photonic reservoir}

\let\comma,

\author{Rosario Di Bartolo}
\affiliation{Dipartimento di Fisica - Sapienza Università di Roma\comma{} P.le Aldo Moro 5\comma{} I-00185 Roma\comma{} Italy}
\author{Valeria Cimini}
\affiliation{Dipartimento di Fisica - Sapienza Università di Roma\comma{} P.le Aldo Moro 5\comma{} I-00185 Roma\comma{} Italy}
\author{Giorgio Minati}
\affiliation{Dipartimento di Fisica - Sapienza Università di Roma\comma{} P.le Aldo Moro 5\comma{} I-00185 Roma\comma{} Italy}
\author{Danilo Zia}
\affiliation{Dipartimento di Fisica - Sapienza Università di Roma\comma{} P.le Aldo Moro 5\comma{} I-00185 Roma\comma{} Italy}
\author{Luca Innocenti}
\affiliation{Universit\`a degli Studi di Palermo\comma{} Dipartimento di Fisica e Chimica - Emilio Segr\`e\comma{} via Archirafi 36\comma{} I-90123 Palermo\comma{} Italy}
\author{Salvatore Lorenzo}
\affiliation{Universit\`a degli Studi di Palermo\comma{} Dipartimento di Fisica e Chimica - Emilio Segr\`e\comma{} via Archirafi 36\comma{} I-90123 Palermo\comma{} Italy}
\author{Gabriele Lo Monaco}
\affiliation{Universit\`a degli Studi di Palermo\comma{} Dipartimento di Fisica e Chimica - Emilio Segr\`e\comma{} via Archirafi 36\comma{} I-90123 Palermo\comma{} Italy}
\author{Nicol\`{o} Spagnolo}
\affiliation{Dipartimento di Fisica - Sapienza Università di Roma\comma{} P.le Aldo Moro 5\comma{} I-00185 Roma\comma{} Italy}
\author{Taira Giordani}
\affiliation{Dipartimento di Fisica - Sapienza Università di Roma\comma{} P.le Aldo Moro 5\comma{} I-00185 Roma\comma{} Italy}
\author{G. Massimo Palma}
\affiliation{Universit\`a degli Studi di Palermo\comma{} Dipartimento di Fisica e Chimica - Emilio Segr\`e\comma{} via Archirafi 36\comma{} I-90123 Palermo\comma{} Italy}
\author{Mauro Paternostro}
\email{mauro.paternostro@unipa.it}
\affiliation{Universit\`a degli Studi di Palermo\comma{} Dipartimento di Fisica e Chimica - Emilio Segr\`e\comma{} via Archirafi 36\comma{} I-90123 Palermo\comma{} Italy}
\affiliation{Centre for Quantum Materials and Technologies, School of Mathematics and Physics, Queen’s University Belfast, BT7 1NN, United Kingdom}
\author{Alessandro Ferraro}
\affiliation{Dipartimento di Fisica Aldo Pontremoli\comma{} Universit\`a degli Studi di Milano\comma{} I-20133 Milano\comma{} Italy}
\author{Fabio Sciarrino }
\email{fabio.sciarrino@uniroma1.it}
\affiliation{Dipartimento di Fisica - Sapienza Università di Roma\comma{} P.le Aldo Moro 5\comma{} I-00185 Roma\comma{} Italy}

\begin{abstract}
Model-independent estimation of the properties of quantum states is a central challenge in quantum technologies, as experimental imperfections, drifts, and imprecise models of the actual quantum dynamics inevitably hinder accurate reconstructions.
Here, we introduce a training strategy for photonic quantum extreme learning machines in which both the learning stage and the optimization of the measurement settings are performed entirely with classical light, while inference is carried out on genuinely quantum states. The protocol is based on the identity between the normalized output intensities following the evolution of 
coherent states through a linear optical reservoir, 
and the output statistics obtained with separable input quantum states.
Building on this correspondence, we implemented a model-free, gradient-based optimization of the reservoir measurement projection directly on experimental data, without relying on a prior model of the device transformation. We experimentally show that the resulting classical-to-quantum transfer enables accurate reconstruction of single-qubit Pauli observables for previously unseen single-photon states, and extends to the estimation of a two-qubit entanglement witness for arbitrary bipartite states. Beyond demonstrating a qualitatively distinct form of out-of-distribution generalization across the classical-to-quantum boundary, our results identify a practical route to fast, adaptive, and resource-efficient training of photonic quantum learning devices.
\end{abstract}

\maketitle

\section*{Introduction}

The accurate experimental reconstruction of quantum states, and their properties, remains a central challenge in quantum information science~\cite{paris2004QuantumStateEstimation,elben2022RandomizedMeasurementToolbox}. Approaches such as full quantum state tomography and shadow-based methods typically rely on the accurate modelling of the underlying quantum dynamics and measurement apparatus~\cite{aaronson2018ShadowTomographyQuantum,huang2020PredictingManyProperties,elben2022RandomizedMeasurementToolbox}. Single-setting estimation strategies have been explored as a potential route to reducing the measurement overhead associated with repeatedly reconfiguring a measurement apparatus to implement projective measurements over various bases~\cite{li2024EstimatingManyProperties,acharya2021ShadowTomographyBased,nguyen2022OptimizingShadowTomography,lomonaco2025NonstabilizernessCostQuantum,stricker2022ExperimentalSingleSettingQuantum,bian2015RealizationSingleQubitPositiveOperatorValued,an2024EfficientCharacterizationsMultiphoton,oren2017QuantumStateTomography,smania2020ExperimentalCertificationInformationally,fischer2022AncillafreeImplementationGeneralized,ivashkov2024HighFidelityMultiqubitGeneralized}, but these approaches also require a reliable characterization of the effective measurement map in order to estimate the properties of interest from measurement data. In experimental platforms, this requirement can become a severe limitation: device imperfections, temporal drifts, and limited characterized noise make accurate modelling difficult and can substantially degrade reconstruction performance.
This motivates the search for estimation strategies that can operate without requiring a detailed model of the internal dynamics of evolution and measurement apparatus.

Quantum reservoir computing (QRC) embodies a promising approach to address these issues.
In QRC, a quantum system with fixed internal dynamics is used to scramble quantum information into a higher-dimensional space, and a simple linear readout is trained to find the best way to extract target properties of the input state from measurement data~\cite{negoro2018MachineLearningControllable,ghosh2019QuantumReservoirProcessing,nakajima2019BoostingComputationalPower,martinez-pena2020InformationProcessingCapacity,fujii2021QuantumReservoirComputing,nokkala2021GaussianStatesContinuousvariable,mujal2021OpportunitiesQuantumReservoir,martinez-pena2021DynamicalPhaseTransitions,mujal2023TimeseriesQuantumReservoir,llodra2023BenchmarkingRoleParticle,kobayashi2024FeedbackDrivenQuantumReservoir,sannia2024DissipationResourceQuantum,palacios2024RoleCoherenceManybody,nokkala2024RetrievingQuantumFeatures,cindrak2026MemoryNonlinearityTradeoffQuantum,prietogarcia2026QuantumReservoirComputing,sannia2025NonMarkovianityMemoryEnhancement,sannia2025ExponentialConcentrationSymmetries,llodra2025QuantumReservoirComputing,das2025QuantumReservoirComputing}. Whereas QRC is in general suited to deal with time series, its memoryless counterparts, referred to as quantum extreme learning machines (QELMs), operate on time-independent data~\cite{innocenti2023potential,lomonaco2024QuantumExtremeLearning,xiong2025FundamentalAspectsQuantum,assil2025EntanglementEstimationWerner,gili2026LearningFunctionsQuantum,assil2026MemoryenhancedQuantumExtreme}.
Experimental demonstrations have recently been reported for both QRC~\cite{garcia-beni2023ScalablePhotonicPlatform,labay-mora2024NeuralNetworksQuantum,senanian2024MicrowaveSignalProcessing,das2025ImageDenoisingQuantum,paparelle2025ExperimentalMemoryControl,dibartoloTimeseriesForecastingMultiphoton2025,hou2026HighAccuracyTemporalPrediction, sakurai2025quantum} and QELM~\cite{suprano2024experimental,zia2025quantum,cimini2025large,joly2025HarnessingPhotonIndistinguishability,brusaschi2026QuantumInferenceClassically,dao2026BreakingConcentrationBarriers,swierczewski2026QuantumReservoirComputing} architectures, remarking the rapid development of reservoir-based quantum learning platforms.
The QELMs approach is especially appealing for estimation tasks: rather than relying on an explicit model of the experimental apparatus, QELMs use supervised learning to directly infer how a set of given measurement outcomes should be processed so as  to estimate quantities of interest from previously unseen data. In this sense, QELM-based estimation strategies fit into the broader scope of self-calibration~\cite{rehacek2010OperationalTomographyFitting,mogilevtsev2012SelfcalibrationSelfconsistentTomography,onorati2024NoisemitigatedRandomizedMeasurements,cooper2014LocalMappingDetector,branczyk2012SelfcalibratingQuantumState,zhang2020ExperimentalSelfCharacterizationQuantum}.

The 
experimental demonstrations in photonic platforms~\cite{suprano2024experimental,zia2025quantum,cimini2025large,brusaschi2026QuantumInferenceClassically} show that QELMs can extract information about input states from noisy and imperfect devices more effectively than alternatives based on shadow tomography operated on the same hardware
~\cite{suprano2024experimental,zia2025quantum}. The reason behind such effectiveness lies in a favorable trade-off in prior knowledge: QELM-based protocols replace detailed modeling of the device dynamics with access to a training set of known input states, while shadow-based ones typically require a reliable characterization of the effective measurement and full experimental transformation. In many practical settings, the latter requirements are more challenging than 
the preparation of a restricted family of simple training states with high fidelity. In particular, Ref.~\cite{zia2025quantum} used a photonic QELM to reconstruct an entanglement witness from experimental data and showed that the model could be trained exclusively on separable states while still generalizing to previously unseen entangled inputs. This provided an experimentally relevant instance of out-of-distribution generalization in quantum learning, in the broader context of ongoing efforts to understand transferability and generalization in quantum machine learning~\cite{zen2020TransferLearningScalability,mari2020TransferLearningHybrid,theresajose2023TransferLearningQuantum,caro2023OutofdistributionGeneralizationLearning,pereira2025OutofDistributionGeneralizationLearning,martinperez2026HybridClassicalQuantumTransfer,banchi2021GeneralizationQuantumMachine,caro2021EncodingdependentGeneralizationBounds,huang2021PowerDataQuantum,caro2022GeneralizationQuantumMachine,peters2023GeneralizationOverfittingQuantum,haug2024GeneralizationQuantumMachine,berberich2024TrainingRobustGeneralizable,rodriguezgrasa2026PACBayesianApproachGeneralizationa}.

Here, we introduce a new training paradigm for photonic QELMs that pushes this idea substantially further. We show that the reservoir can be trained entirely with classical light, without loss of performance when the resulting model is later applied to genuinely quantum states. This constitutes a qualitatively distinct form of out-of-distribution generalization. Whereas previous work established generalization across different classes of quantum states --- from separable states during training to entangled states during testing~\cite{zia2025quantum} --- here the transfer occurs across the classical-to-quantum boundary: training is performed with classical coherent light, whereas inference targets single- and two-photon states. 

This possibility is especially powerful and straightforward for photonic platforms, where coherent and nonclassical states can be easily injected into the same physical reservoir, whether realized in bulk optics, fiber-based architectures, or integrated photonic circuits, without modifying the underlying physical transformation.
As a result, the training carried out with bright classical light remains directly transferable to the quantum regime, making photonic platforms exceptionally favourable for classical-to-quantum transfer learning. 
Beyond their conceptual significance, these results have immediate practical implications. Training with classical light dramatically simplifies the experimental workflow and enables faster and more efficient data acquisition. To demonstrate the practical value of this feature, we leverage it to optimize the measurement settings online through a gradient-descent-based procedure that does not require prior knowledge of the underlying system transformation. Crucially, performing the same optimization directly with single-photon states would be experimentally unsuitable, because the long acquisition times in the quantum regime would compromise the stability and effectiveness of the online learning loop. The transfer across the classical-to-quantum boundary therefore does more than demonstrate generalization: it is the enabling mechanism that makes adaptive, online, and fully model-free optimization of a quantum photonic reservoir experimentally viable. Using this strategy, we achieve high-performance reconstruction of single-qubit Pauli observables and of a two-qubit entanglement witness on genuinely quantum states.

Our protocol thus preserves the central advantage of the QELM approach, namely, model-independent learning from the actual experimental device, while rendering the training stage substantially more efficient. 
More broadly, this work expands the scope of transfer learning and domain adaptation in quantum machine learning, identifying a concrete route towards resource-efficient training strategies for photonic quantum learning platforms. In this sense, it also contributes to the broader programme of identifying practically relevant advantages of quantum learning architectures in experimentally realistic settings~\cite{molteni2026ExponentialQuantumAdvantages}.



\section*{Results}

\subsection*{Experimental Setup}

\begin{figure}[t!]
    \centering \includegraphics[width=0.99\linewidth]{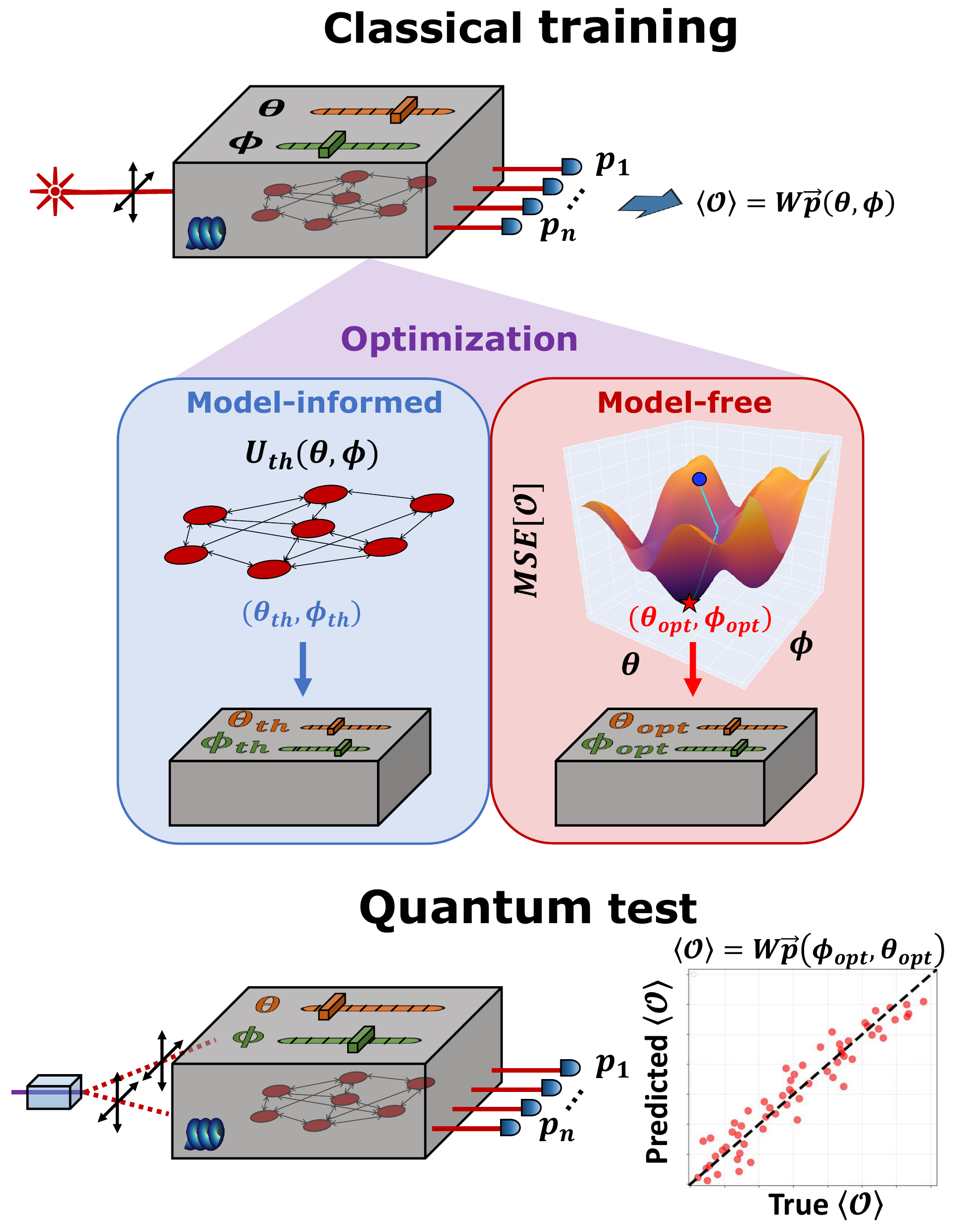}
    \caption{
    \textbf{Classical-to-quantum transfer learning with measurement optimization.} The quantum extreme learning machine (QELM) architecture is trained and optimized using polarization input states generated by a continuous wave (CW) laser. The polarization measurement projections of the reservoir, defined by the pair of angles $(\theta, \phi)$ can be optimized with respect to the target observable $\expval{\mathcal{O}}$. This optimization can be performed either by having knowledge of the reservoir transformation (model-informed) or by adopting a model-free approach based on a gradient descent strategy with feedback, which minimizes the mean square error (MSE) between the predicted and expected values over a fixed training set. The resulting trained model is finally tested on quantum states, consisting of heralded single photons generated through spontaneous parametric down conversion (SPDC) in a nonlinear crystal.}
    \label{fig:scheme}
\end{figure}

Experimental imperfections are a central challenge for the practical deployment of quantum learning architectures on current hardware platforms. Photonic systems offer a distinctive opportunity in this respect, because the same processor can be operated with both classical and quantum light under closely matched conditions. To address this limitation, we adopt a transfer learning strategy in which fast and efficient training is performed using classical light, while the optimized model is subsequently tested on quantum states. This approach is enabled by the equivalence between the normalized output intensities of coherent states and the output statistics obtained with separable quantum states (see Methods for further details), and is implemented within the quantum reservoir computing framework, whose dynamics naturally incorporate experimental imperfections rather than requiring precise control of the underlying transformation.
Importantly, the use of classical light has two distinct roles in our protocol. First, it enables fast and resource-efficient training of the QELM, independently of whether the measurement settings are further optimized. Second, it allows the readout to be refined directly on the experimental platform through an additional optimization step, which can be carried out either using prior knowledge of the system dynamics or through a fully model-free gradient-descent procedure. In this sense, full characterization and fine-tuning of the setup are not essential for the applicability of the protocol, but they can provide an improvement in the reconstruction accuracy of the observable of interest.
A schematic representation of the classical-to-quantum transfer learning protocol within the QELM framework is illustrated in Fig. \ref{fig:scheme}.

\begin{figure}[htb!]
    \centering \includegraphics[width=0.99\linewidth]{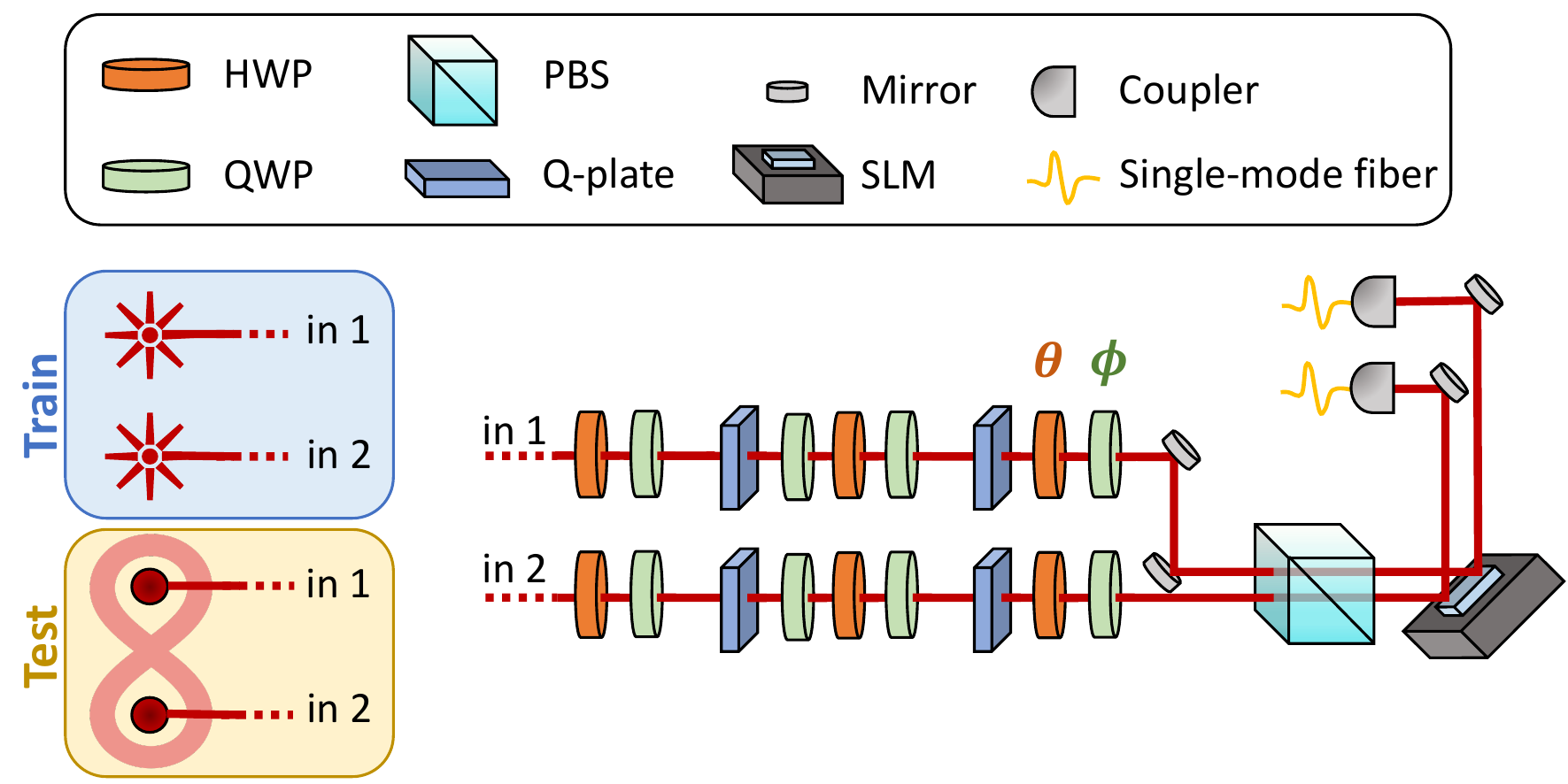}
    \caption{\textbf{Experimental platform.} Laser light and photon pairs at $\lambda = 808$ $nm$ provide the classical training and quantum test sets of the QELM architecture, respectively. The elements of these datasets are encoded in the polarization degree of freedom of the input states, through a half-wave plate (HWP) followed by a quarter-wave plate (QWP). A series of q-plates and waveplates in a cascaded configuration are used to implement a two-step QW evolution in the orbital angular momentum (OAM) degree of freedom. Through the coupling between polarization and OAM, the QW acts as the reservoir dynamics of the QELM protocol and enlarges the dimension of the Hilbert space and spreads the input information. The output states are then projected in the polarization space by means of a HWP and QWP followed by a polarizing beam splitter (PBS), with the projection angles $(\theta,\phi)$ iteratively optimized in the model-free strategy. Finally, the distribution of states over the OAM basis $\{\ket{m}, m=-2,-1,0,1,2\}$ is estimated via projective measurement realized with a spatial light modulator (SLM) followed by the coupling to single-mode fibers. The occupation probability is then retrieved and used as the reservoir response to train and test the QELM model.}
    \label{fig:setup}
\end{figure}

The experimental platform realizing the quantum reservoir dynamics and the transfer learning protocol is shown in Fig. \ref{fig:setup}. Classical or quantum signals enter the setup through single-mode fibers (SMFs), then a combination of a half-wave plates (HWPs) followed by a quarter-wave plates (QWPs) is used to modify their polarization and encode the protocol input states. In particular, a continuous wave (CW) laser is used to create a classical training set, while single-photon states produced via spontaneous parametric down conversion (SPDC) in a nonlinear periodically-poled potassium titanyl phosphate (PPKTP) crystal constitute the quantum test set. The input states are then sent through a series of waveplates and q-plates \cite{marrucci2006optical, marrucci2011spin}, which realize two steps of a quantum walk (QW) evolution in the angular momentum degree of freedom of light \cite{giordani2019experimental, suprano2024experimental, zia2025quantum}. Here, leveraging on the conditional evolution of the walker state, the information initially encoded in the bidimensional polarization space is directly transmitted to the larger orbital angular momentum (OAM) one. At the end of the QW platform, the output states are projected in the polarization space by a sequence of HWP, QWP and polarizing beam splitter (PBS) and the occupation probabilities of the state on the OAM basis $\{\ket{m}, m=-2,-1,0,1,2\}$, where $m$ is the eigenvalue of the OAM \cite{allen1992orbital}, are retrieved via projective measurements through a spatial light modulator (SLM) followed by coupling into single-mode fibers. The reconstructed distributions constitute the response of the reservoir, on which the QELM output layer is trained to predict the expectation value of the observable $\mathcal{O}$ under investigation. It is worth noticing that the Hilbert space dimension enlargement allows us to retrieve the input polarization state in a single setting fashion, by measuring only the OAM occupation probability of the QW output states.

Within the QELM framework, no explicit characterization of the effective positive operator-valued measure (POVM) $\mu$, which incorporates both the reservoir evolution and the projective measurement, is required, although the estimation performance depends on its properties, such as its ability to scramble initial information  \cite{innocenti2023potential,vetrano2025StateEstimationQuantum} (see Methods for further details). In our setup, we focused on the polarization projection angles of the final measurement stage, determined by the HWP and QWP angles $\theta$ and $\phi$, respectively (see Fig. \ref{fig:setup}). These measurement parameters can indeed affect the number of resources necessary to produce accurate predictions \cite{suprano2024experimental, zia2025quantum}, i.e. the statistics the QELM needs for calibrating to the experimental setup, and consequently the final estimation precision.
Motivated by this, we adopt a model-free optimization strategy for the measurement projection, allowing the optimal settings to be identified online during the classical training step from the actual system dynamics rather than from prior theoretical assumptions.
We study two complementary regimes. First, we consider the reconstruction of single-photon observables. In this case, only one photon of the generated pairs is injected into the reservoir and used as input state, while the second photon serves as a herald for coincidence detection. Then, we move to study the reconstruction of a two-photon observable, namely an entanglement witness. In this configuration, entangled photon pairs are prepared and both photons are injected into the double quantum-walk architecture.

Specifically, we optimize the photonic QELM for the estimation of the expectation value of a target observable, such as the Pauli operator $\sigma_y$, by iteratively varying the projection waveplate angles and retraining the model to minimize the mean square error (MSE) of its predictions over a fixed dataset. This allows us to build an assumption-free model which intrinsically incorporates experimental imperfections and noise \cite{suprano2021dynamical}. 
Crucially, the ability to train with classical states and infer observables of quantum states \cite{zia2025quantum} unlocks the possibility of optimizing online the reservoir itself on actual experimental outcomes.
Since the optimization can be performed entirely with bright classical light and then transferred to the quantum regime, we efficiently refine the measurement settings in real time, before applying the trained model to single- and two-photon states. Such a strategy would be practically unattainable using quantum states alone, because the acquisition times in the single-photon regime are too long to sustain an effective iterative optimization loop. For further details see Supplementary Information S1.


\subsection*{Gradient-based reservoir optimization}

To optimize the performance of the QELM paradigm, previous works \cite{suprano2024experimental,zia2025quantum} relied on detailed theoretical modeling of the quantum walk dynamics to identify near-optimal measurement settings. Here, we instead adopt a data-driven strategy, in which the effective POVM $\mu$ is directly optimized from measurement data.

We follow a task-dependent approach and minimize the MSE between predicted and exact expectation values of a target observables $\mathcal{O}$. The optimization is performed via gradient descent over a fixed training set, acquired using classical light, a choice that is both resource-efficient and central to the feasibility of the approach. Classical coherent light offers two critical practical advantages during gradient estimation. First, measurements are not shot-noise limited, so the loss function $\mathcal{L}$ can be evaluated with high statistical accuracy at low experimental cost. Second, data acquisition is fast, which minimizes the risk of optical misalignment between the two evaluations of $\mathcal{L}$ required by the finite-difference estimator, a source of systematic error that would otherwise corrupt the gradient. Together, these properties allow reliable exploration of the parameter landscape and stable convergence (see the Supplementary Information S2 for further details on the experimental landscape). At each step, the chosen parameters of the reservoir are updated according to a feedback mechanism so as to minimize  the MSE. 

\begin{figure*}[htb!]
    \centering \includegraphics[width=1\linewidth]{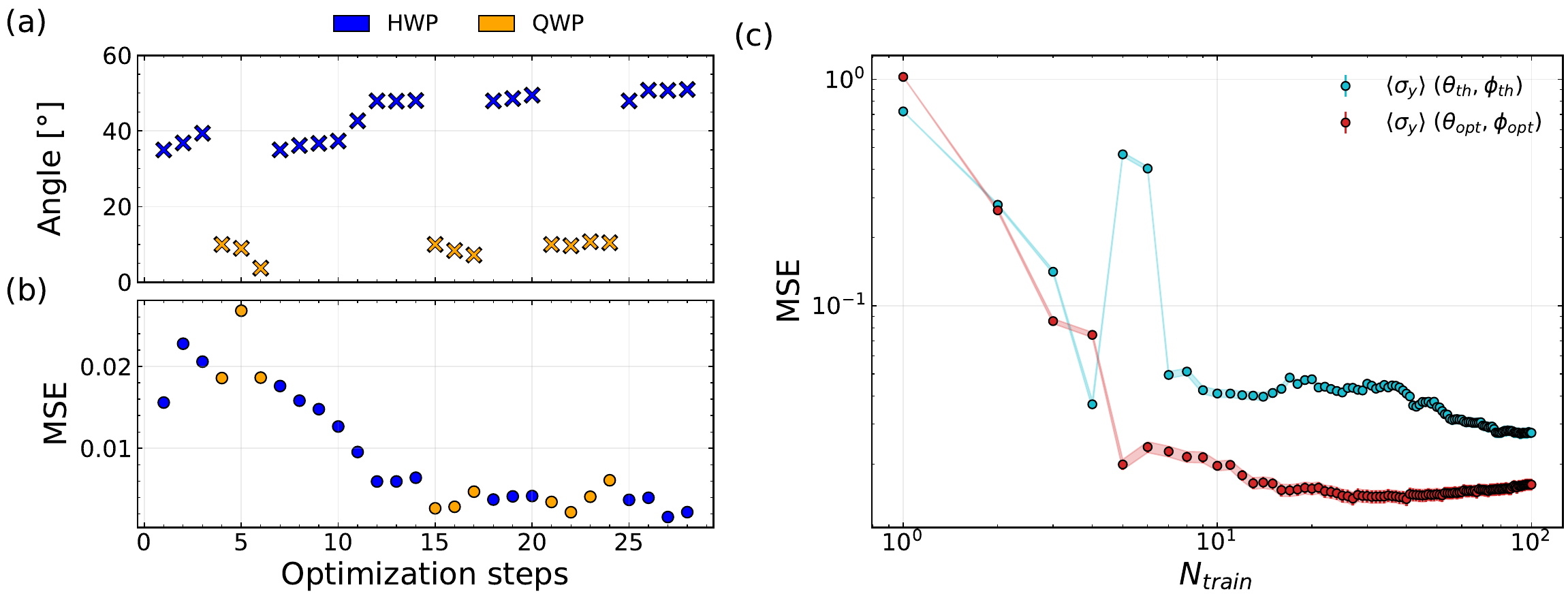}
    \caption{
    \textbf{Gradient descent reservoir optimization and performance for $\langle \sigma_y \rangle$ reconstruction.} (a) Evolution of the wave plate angles during the optimization step. The HWP (blue) and QWP (orange) angles are updated alternately: at each step, one plate configuration is optimized while keeping the angle of the other fixed to its current best value. (b) The MSE over the training dataset is evaluated at each optimization step and used as the loss function. At each step, 15 randomly coherent light input states are prepared in polarization, and the output probability distribution in the orbital angular momentum resulting from the single-photon two-step quantum walk is measured and used to learn the expectation value $\langle \sigma_y \rangle$. (c) MSE over the test set, consisting of $100$ single-photon polarization states, as a function of the number of training examples taken from a set of 100 randomly generated coherent input polarization states. The performance obtained with measurement settings derived from the theoretical model of the photonic platform ($\theta_{th}$, $\phi_{th}$) (blue), is compared with the one obtained using experimental optimized settings ($\theta_{opt}$, $\phi_{opt}$) (red). Uncertainties on the curves are estimated through 100 Monte Carlo realizations, where photon-count distributions are resampled assuming Poisson statistics. Shaded regions denote one standard deviation.
    \label{fig:1}}
\end{figure*}

In our implementation, each MSE evaluation during the gradient descent optimization is computed over $15$ randomly sampled input states, so as to provide a sufficiently expressive training set for generalization. We optimize the projection stage of the quantum walk, corresponding to the polarization measurement defined by the HWP and QWP with angles $\theta$ and $\phi$, respectively (see Fig. \ref{fig:setup}). To reduce the number of gradient evaluations, and consequently the experimental overhead, we employ an alternating optimization scheme in which only one parameter is updated at a time while keeping the other fixed. The update rules are given by
\begin{equation}
\nu^{(i+1)}_k = \nu^{(i)}_k - \eta \,\nabla_{\nu_k} \mathcal{L}\!\left(\nu^{(i)}_k, \nu^{(i)}_{j\neq k}=\nu^*_{j\neq k}\right),
\end{equation}
where $\eta$ is the learning rate, ${\bm \nu}=(\theta,\phi)$ is the vector of parameters and starred quantities stand for their respective temporarily fixed values. The gradients are estimated via finite differences
\begin{equation}
\nabla_{\nu_k} \mathcal{L}\!\left(\nu^{(i)}_k, \nu_{j\neq k}^{*}\right) 
= \frac{\mathcal{L}(\nu^{(i)}_k + \epsilon, \nu^{*}_{j\neq k}) - \mathcal{L}(\nu^{(i)}_k - \epsilon, \nu^{*}_{j\neq k})}{2 \epsilon},
\end{equation}
where $\epsilon$ is a small increment.
Finally, each parameter is updated iteratively until convergence or until the MSE ceases to improve. The parameter is then fixed at the value yielding the lowest loss, and the optimization proceeds with the other parameter. This alternating procedure is repeated until convergence, providing an experimentally efficient coordinate-descent strategy for tailoring the quantum feature map to the task. The experimental waveplate settings, together with the MSE evaluation for the chosen $\sigma_y$ Pauli observable, are reported in Fig. \ref{fig:1}(a)-(b). 
The optimal experimental parameters $\theta_{\rm opt}$ and $\phi_{\rm opt}$, found with such optimization strategy, are then used for the quantum inference stage, where testing is performed by acquiring a dataset of single-photon states. For further details see Supplementary Information S1.

\subsection*{QELM performances after training with classical light}

To benchmark the effectiveness of this approach, we start by studying the model performance on single-photon Pauli observables reconstruction. For this purpose, we collect $100$ random classical training states, which are injected into the photonic platform and evolved through the single quantum walk, acting as the reservoir of the QELM. 
We compare two different configurations for the output measurement projections. In the first case, the projection settings are determined through the gradient descent optimization described above, evaluated directly on experimental outcomes. In the second case, instead, the projection angles are chosen from the knowledge of the QW dynamics, thus selecting hidden-layer configurations that provide an approximately uniform coverage of the OAM space. 
This comparison allows us to directly evaluate the advantage of a fully data-driven optimization over a strategy based on the theoretical knowledge of the reservoir dynamics, which do not incorporate the actual noise present in the experimental setup.

\begin{figure*}[htbp]
    \centering \includegraphics[width=0.99\linewidth]{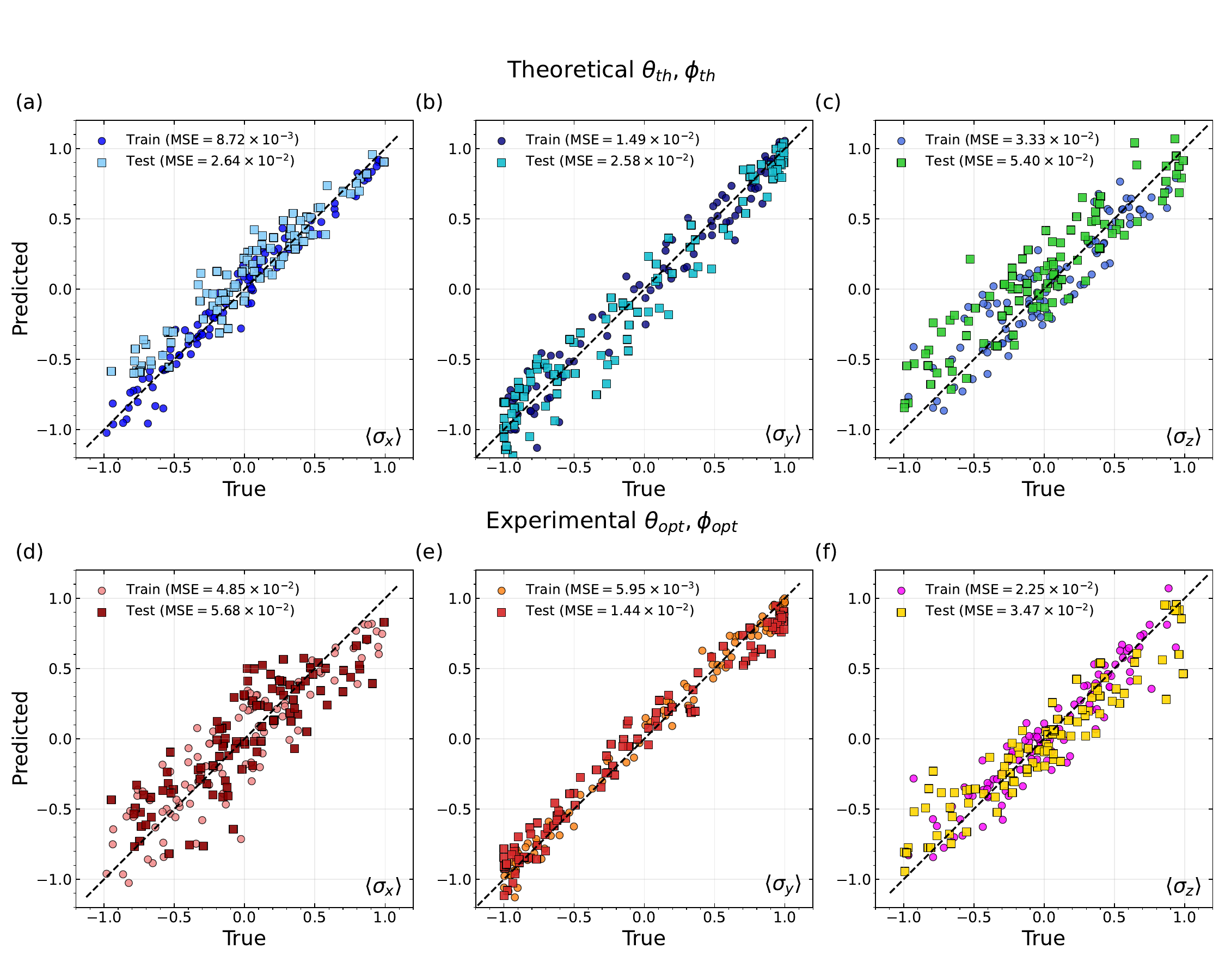}
    \caption{
    \textbf{Experimental reconstruction of Pauli observables via transfer learning.} QELM predicted versus true values of the Pauli expectation values $\langle \sigma_x \rangle$, $\langle \sigma_y \rangle$,  and $\langle \sigma_z \rangle$. The dataset consists of 100 random coherent light input polarization states, used for training the model, and 100 random single-photon input polarization states for its testing. (a), (b), (c) reports the training and test predictions using the reservoir with the theoretical angles ($\theta_{th}$, $\phi_{th}$). (d), (e), (f) reports the training and test predictions using the reservoir with the experimentally optimized angles ($\theta_{opt}$, $\phi_{opt}$) found optimizing the reconstruction of $\langle \sigma_y \rangle$. The MSE values of the training and test sets are reported in the legend.
    \label{fig:3}}
\end{figure*}

Having optimized the reservoir with classical data, we then investigate whether the resulting trained models can be successfully transferred to the quantum domain. To this end, we use the classical training sets obtained in the two measurement configurations described above and evaluate their performance in reconstructing observables for single-photon states. In particular, we prepare a test set of $100$ single-photon input states and assess the QELM in estimating the expectation value of the Pauli operator $\langle \sigma_y \rangle$ using models trained exclusively on classical light states. The performance on the test set, as a function of the number of classical training examples, is reported in Fig. \ref{fig:1}(c), validating the transfer capability of the model.
The experimental results show that optimization performed directly on measured outcomes consistently leads to a lower MSE. As the number of training examples increases, both approaches exhibit an overall decrease in the test error, indicating improved generalization with larger training datasets. However, the experimentally optimized measurement settings $(\theta_{opt}, \phi_{opt})$ systematically outperform the theoretically derived ones $(\theta_{th}, \phi_{th})$ across the entire training range. While the model-informed strategy initially follows a similar trend, its performance saturates at a higher MSE and displays larger fluctuations for intermediate training-set sizes, highlighting its sensitivity to discrepancies between the theoretical description and the actual experimental conditions.
This improvement can be interpreted as a direct consequence of the data-driven procedure capturing the true response of the experimental apparatus, including nonideal effects that are inevitably neglected in an ideal QW-based design. These results highlight the practical value of the proposed method and demonstrate that adaptive optimization of the measurement settings can significantly enhance the protocol performance. To test the robustness of the optimized configuration beyond the target observable, we also evaluated the reconstruction of the remaining Pauli observables using the same experimentally optimized settings $(\theta_{opt}, \phi_{opt})$ obtained for $\langle \sigma_y \rangle$ (see Fig.\ref{fig:3}). Further analysis, including the performance as a function of the training set size, for the non-optimized observables, is discussed in Supplementary Information S2.

\begin{figure}[htbp]
    \centering \includegraphics[width=0.99\linewidth]{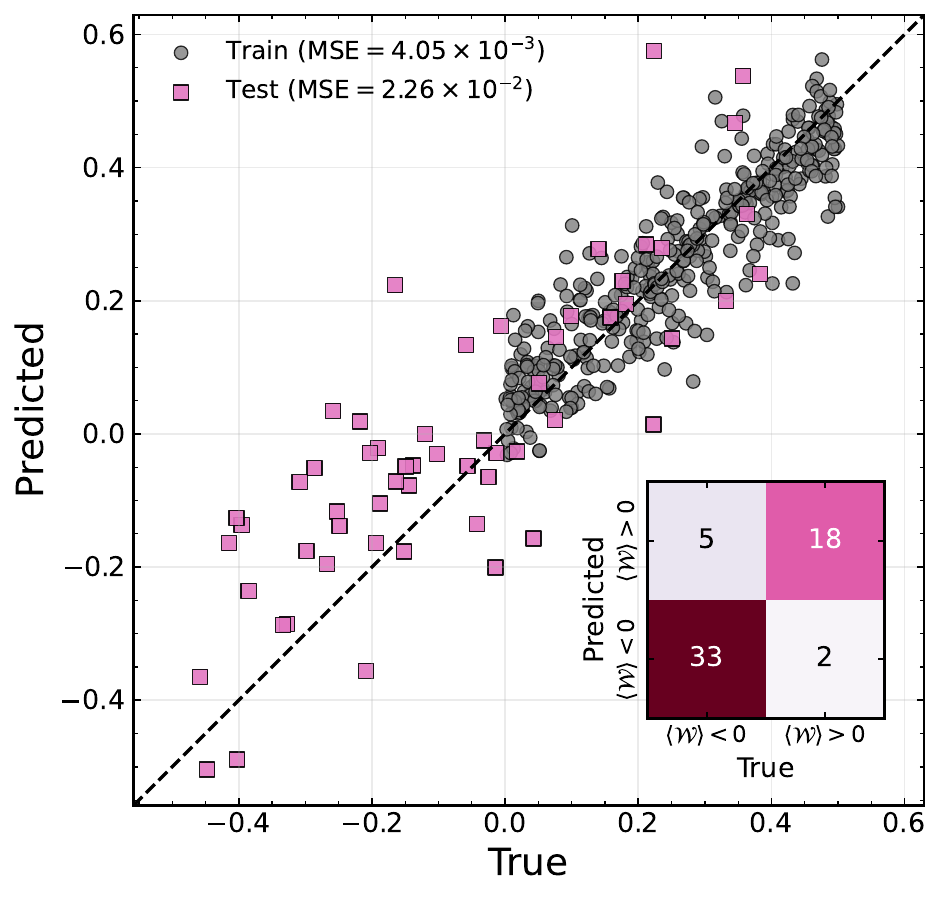}
    \caption{
    \textbf{Experimental entanglement witness reconstruction via transfer learning.}
    Predicted versus true values of the entanglement witness $\langle \mathcal{W} \rangle$ for a QELM trained on $400$ classical polarization states (grey), prepared by applying random local polarization rotations to horizontally polarized light, and tested on $58$ randomly prepared two-photon entangled states (pink), obtained by applying random local polarization rotations to the maximally entangled Bell state $\lvert \Psi^{-} \rangle = \frac{1}{\sqrt{2}} ( \lvert HV \rangle - \lvert VH \rangle )$. The MSE values for both sets are reported in the legend. The lower-right corner shows the confusion matrix of the test set, whose diagonal entries indicate the number of correctly classified states, while the off-diagonal entries correspond to false positives and false negatives.
    \label{fig:4}}
\end{figure}

Further evidence of the classical-to-quantum transfer capability is provided by the reconstruction of different Pauli observables. We find that the model employing gradient descent-optimized measurement settings achieves the best performance for the observable used during optimization [Fig. \ref{fig:3}(b) and (e)], while still maintaining high accuracy in the estimation of the remaining Pauli operators. This behavior indicates that the learned measurement settings capture general features of the reservoir response that remain informative across different observables.


Finally, we study the reconstruction of a two-photon observable, namely the entanglement witness $\mathcal{W}$.
This represents the most relevant test of the proposed approach. While the reconstruction of single-photon Pauli observables already demonstrates the transfer learning capability of the QELM approach from classical training data to quantum inputs, the true potential is demonstrated when reconstructing intrinsically quantum features, while still trained on only classical inputs.
For this reason, we focus on the estimation of an entanglement witness, whose expectation value directly probes nonclassical correlations between two photons.
We choose as target observable an entanglement witness whose value is positive for all separable states, while negative values certify the presence of entanglement \cite{guhne2009entanglement}. Specifically, we consider the witness $\mathcal{W} = \frac{1}{2} ( I - \lvert \Psi^{+} \rangle \langle \Psi^{+} \rvert )$, where $I$ is the identity operator and $\vert \Psi^{+} \rangle = \frac{1}{\sqrt{2}}(\vert HV\rangle + \vert VH\rangle)$ is a maximally entangled Bell state.
For this study, 
starting from horizontally polarized coherent light, we prepare $400$ different classical polarization input states by applying local polarization rotations via the preparation HWP and QWP, and perform the training stage.  In this configuration, laser light is injected into both quantum walks, and the resulting measurement outcomes are used to train the model. The trained QELM is then tested on the outcomes obtained from $58$ two-photon entangled states, prepared by first generating $\lvert \Psi^{-} \rangle = \frac{1}{\sqrt{2}}(|HV\rangle - |VH\rangle)$, applying random local polarization rotations, and injecting the resulting states into the reservoir.
As shown in Fig. \ref{fig:4}, the resulting MSE confirms that, although trained solely with classical light, the model is still able to reconstruct witness expectation values below zero, thus correctly revealing the presence of entanglement.
This demonstrates that the learned mapping is able to capture nonclassical features that are never explicitly present in the training data. In other words, the classical-to-quantum transfer does not merely preserve performance on simple local observables, but extends to genuinely multipartite quantum properties such as entanglement detection. Further analysis on the robustness of the transfer learning approach is discussed in Supplementary Information S2.

\section*{Discussion}

We have introduced and experimentally demonstrated a training paradigm for photonic QELMs in which the reservoir is trained and optimized exclusively with classical light and the learned model can seamlessly estimate properties of previously unseen single- and two-photon quantum states, thus establishing a form of out-of-distribution generalization across the classical-to-quantum boundary.
This generalization is achieved without an explicit model of the reservoir dynamics or of the physical measurement performed by the apparatus. In this sense, the protocol retains the main advantage of the QELM-based estimation, namely, learning directly from the actual experimental apparatus, while making the training stage substantially faster and more practical.

A key aspect of the method is that classical coherent light makes it possible to evaluate the loss function rapidly and with high statistical precision, thereby enabling online optimization of the measurement projection through a fully data-driven feedback loop. This optimization would be extremely demanding in the single-photon regime, where long acquisition times would not allow for collecting enough statistics to implement such a strategy. By shifting the training stage to the classical domain, we can instead tailor the reservoir response directly from measured data and then transfer the resulting configuration to quantum inputs. The experimental comparison with measurement settings chosen from prior theoretical modeling shows that this strategy yields lower reconstruction error, indicating that learning directly on the device response captures features that are not fully accounted for by an idealized description of the setup.

The transfer is not limited to the reconstruction of the observable used during optimization. After optimizing the reservoir to estimate $\langle \sigma_y \rangle$, we observe that the same measurement configuration still provides accurate predictions for the other single-qubit Pauli observables on previously unseen single-photon states. This suggests that the learned feature map is not narrowly specialized to a single target, but preserves broader task-relevant information about the input state. More importantly, the transfer extends beyond local observables: a model trained only on classical states is able to predict negative values of a two-qubit entanglement witness for Bell states, thereby correctly identifying entanglement even though no genuinely quantum correlations are present in the training set.

Taken together, these results establish a form of out-of-distribution generalization across the classical-to-quantum boundary in an experimentally realistic photonic platform. More broadly, they indicate that the favourable trade-off underlying QELM-based estimation can be pushed further than previously shown: detailed device characterization can be replaced not only by supervised training on simple quantum states, but even by training on classical optical states. 
Crucially, this opportunity is especially meaningful for photonic platforms where the same hardware can be operated with classical and quantum light.
This allows the learning and optimization stage to be performed rapidly and efficiently with bright fields, while preserving direct applicability to the quantum regime, making this transfer not only a new form of generalization, but also a distinctive operational advantage of photonic learning architectures.
This opens a concrete route towards adaptive and resource-efficient quantum photonic learning protocols, and suggests new directions for studying domain adaptation and generalization in quantum machine learning.

\section*{Materials and Methods}

\subsection*{State estimation via QELMs}

Quantum reservoir computing uses a quantum system as a reservoir for processing time-dependent data. A defining feature of QRC is the presence of memory: the output produced at a given step generally depends not only on the current input but also on previous ones. Quantum extreme learning machines are the memoryless counterpart of this paradigm~\cite{innocenti2023potential,suprano2024experimental,zia2025quantum}. In a QELM, each input state is processed independently by a fixed quantum device, and only a linear readout acting on the measured probabilities is trained.

In this setting, the reservoir is modelled as a fixed quantum channel $\Phi$ followed by a physical POVM $\mu^{\mathrm{phys}}=\{\mu_b^{\mathrm{phys}}\}_{b=1}^{n_{\mathrm{out}}}$. Crucially, neither $\Phi$ nor $\mu^{\mathrm{phys}}$ needs to be known explicitly during training: the device is treated as a black box whose input-output statistics are learned directly from data. Given an input state $\rho\in\mathcal{H}_{\mathrm{in}}$, the corresponding outcome probabilities are $p_b(\rho)=\trace[\mu_b^{\mathrm{phys}}\,\Phi(\rho)]$.
We collect them into the probability vector $\mathbf{p}(\rho)=(p_1(\rho),\dots,p_{n_{\mathrm{out}}}(\rho))^{T}\in\mathbb{R}^{n_{\mathrm{out}}}$. Because both the channel and the trace are linear, the map $\rho\mapsto\mathbf{p}(\rho)$ is linear. It is therefore convenient to absorb the reservoir dynamics and the physical measurement into a single effective POVM on the input space,
namely $\mu_b=\Phi^{\dagger}(\mu_b^{\mathrm{phys}})$, where $\Phi^{\dagger}$ denotes the adjoint of the channel. In this notation, $p_b(\rho)=\trace[\mu_b\rho]$.
The effective POVM, rather than $\Phi$ and $\mu^{\mathrm{phys}}$ separately, is the object that directly determines the accessible information and hence the estimation performance~\cite{innocenti2023potential}.

Suppose now that the goal is to estimate a family of $\nobs$ observables $\bscalO=\{\calO_j\}_{j=1}^{\nobs}$ from measurement data, using a training set of $\ntr$ known input states $\{\rho_i\}_{i=1}^{\ntr}$. The corresponding target matrix is $Y_{\bscalO}\in\mathbb{R}^{\nobs\times\ntr}$, with entries $(Y_{\bscalO})_{ji}=\trace[\calO_j\rho_i]$, while the probability matrix is $P\in\mathbb{R}^{n_{\mathrm{out}}\times n_{\mathrm{tr}}}$, with entries $P_{bi}=p_b(\rho_i)=\trace[\mu_b\rho_i]$. The readout is described by a matrix $W_{\bscalO}\in\mathbb{R}^{n_{\mathrm{obs}}\times n_{\mathrm{out}}}$ that maps measured probabilities to estimated expectation values via $\hat{\mathbf{y}}(\rho)=W_{\bscalO}\mathbf{p}(\rho)$. In the ideal infinite-statistics limit, training amounts to solving the linear system $Y_{\bscalO}=W_{\bscalO}P$.
A standard choice is ridgeless linear regression,
\begin{equation}\label{eq:W_as_argmin}
    W_\bscalO=\operatorname*{arg\,min}_{X\in\mathbb{R}^{n_{\mathrm{obs}}\times n_{\mathrm{out}}}}\|XP-Y_{\bscalO}\|_{2}^{2},
\end{equation}
which has canonical solution $W_\bscalO=Y_{\bscalO}P^{+}$, where $P^{+}$ denotes the Moore-Penrose pseudoinverse.
No bias term is included in Eq. \ref{eq:W_as_argmin}, because expectation values of observables depend linearly, rather than affinely, on the input density matrix.

In practice, the exact probabilities are not directly accessible and must be estimated from finite measurement statistics. We denote by $\hat P_N$ the matrix whose $i$-th column contains the empirical frequency vector obtained from $N$ measurement shots on the training state $\rho_i$. Replacing $P$ with $\hat P_N$ yields the finite-statistics estimate $\hat W_{\bscalO,N}=Y_{\bscalO}\hat P_{N}^{+}$.
Finite training statistics therefore introduce sampling noise in the features, which propagates nonlinearly through the pseudoinverse and can substantially affect the learned readout~\cite{innocenti2023potential}.
Consider now the testing stage. If a previously unseen test state $\sigma$ is measured $N_{\mathrm{test}}$ times, one obtains an empirical probability vector $\hat{\mathbf{p}}_{N_{\mathrm{test}}}(\sigma)$ and the predicted observables become $\hat{\mathbf{y}}(\sigma)=\hat W_{\bscalO,N}\hat{\mathbf{p}}_{N_{\mathrm{test}}}(\sigma)$. Hence training and test statistics play conceptually distinct roles: $N$ determines how accurately the readout matrix $W_\bscalO$ is learned, while $N_{\mathrm{test}}$ determines the statistical uncertainty of the final prediction for a previously unseen state. In the large-statistics limit, $\hat P_N\to P$ and $\hat{\mathbf{p}}_{N_{\mathrm{test}}}(\sigma)\to\mathbf{p}(\sigma)$, albeit $\hat P_N^+\to P^+$ might fail in some regimes.
The overall 
estimation performances depends on the information-processing properties of $\mu$: the reservoir must encode input-state information into a sufficiently rich probability vector while preserving task-relevant structure~\cite{suprano2024experimental,zia2025quantum,vetrano2025StateEstimationQuantum}. The readout then selects the relevant linear combination of outcomes. Crucially, because the mapping from measured probabilities to target observables is learned directly from input-output pairs, accurate reconstruction does not require a full physical model of imperfections and noise, which is the key model-free advantage of the approach~\cite{suprano2021dynamical,suprano2024experimental}.


\subsection*{Coherent vs quantum evolution}

Let us now specialize the general formalism to the case in which the input information is encoded in polarization qubits and the reservoir is a linear optical device. The aim of this section is to clarify the conceptual difference between quantum and coherent inputs, focusing on how the input state is mapped to output probabilities or intensities in the single- and two-photon cases, as compared with coherent-state inputs.

\paragraph{Quantum inputs, one spatial mode}
Consider as input a single spatial mode with two polarization modes, $H$ and $V$. In the single-photon case, the input qubit is
\begin{equation}
    \ket{\psi_{\rm in}} =
    c_H \ket{H}+c_V \ket{V} =
    \left(c_H a_H^\dagger + c_V a_V^\dagger\right)\ket{0},
\end{equation}
with $|c_H|^2+|c_V|^2=1$.
The linear optical apparatus acts on the creation operators as $a_\mu^\dagger \to \sum_{m=1}^{L} U_{m,\mu}\, b_m^\dagger$, with $\mu\in\{H,V\}$,
where $m$ labels the output OAM modes of the device. The output state is therefore
\begin{equation}
    \ket{\psi_{\rm out}} =
    \sum_{m=1}^{L} \left(U_{m,H}c_H+U_{m,V}c_V\right)b_m^\dagger\ket{0}.
\end{equation}
Hence the output probability of detecting a photon in the mode $m$ is
\begin{equation}\label{eq:single_photon_probability}
    p_m^{(1)} = \left|U_{m,H}c_H+U_{m,V}c_V\right|^2
    = |(Uc)_m|^2,
\end{equation}
writing $c=(c_H,c_V)^T$.

\paragraph{Quantum inputs, two spatial modes}
Consider now as input two spatial modes, labelled $1$ and $2$, each carrying a polarization qubit, with a single photon each. Label the input modes as $1H,1V,2H,2V$. A general two-qubit input state is then written as
\[
\ket{\Psi_{\rm in}}
=
\sum_{\mu,\nu\in\{H,V\}} c_{\mu\nu}\,
a_{1\mu}^\dagger a_{2\nu}^\dagger \ket{0},
\quad
\sum_{\mu,\nu}|c_{\mu\nu}|^2=1.
\]
The linear optical apparatus again acts as
$a_{s\mu}^\dagger \to \sum_{m=1}^{L} U_{m,s\mu}\, b_m^\dagger$, where $s\in\{1,2\}$ labels the spatial input modes and $\mu\in\{H,V\}$ the polarization modes. Therefore,
\[
\ket{\Psi_{\rm out}}
=
\sum_{m,n=1}^{L}
\left(
\sum_{\mu,\nu} c_{\mu\nu}\, U_{m,1\mu}U_{n,2\nu}
\right)
b_m^\dagger b_n^\dagger \ket{0}.
\]
For two distinct output modes $m\neq n$, the coincidence probability is
\[
p_{mn}^{(2)}
=
\left|
\sum_{\mu,\nu}
c_{\mu\nu}
\left(
U_{m,1\mu}U_{n,2\nu}
+
U_{m,2\mu}U_{n,1\nu}
\right)
\right|^2.
\]
In the special case where the two-qubit input is a product state, the coefficients factorize as
$c_{\mu\nu}=c_\mu^{(1)}c_\nu^{(2)}$, where $c^{(1)}$ and $c^{(2)}$ are the polarization vectors of the two input photons. Note that even in this case the output state is not a product state, because because the unitary $U$ generally couples the two spatial modes.

\paragraph{Coherent input: single spatial mode.}
Consider now the same linear optical apparatus, but with a coherent input instead of a single photon. For one spatial mode, the input is a coherent state in the two polarization modes,
\[
\ket{\alpha_H,\alpha_V}
=
\ket{\alpha_H}_H\otimes\ket{\alpha_V}_V.
\]
To encode the same polarization qubit as above, we set $\alpha_H=\alpha\, c_H$,
$\alpha_V=\alpha\, c_V$,
with $\alpha\in\mathbb C$ the overall field amplitude. The logical input is therefore carried by the \emph{normalized polarization vector}
\[
\frac{1}{\sqrt{|\alpha_H|^2+|\alpha_V|^2}}
\begin{pmatrix}
\alpha_H\\
\alpha_V
\end{pmatrix}
=
\begin{pmatrix}
c_H\\
c_V
\end{pmatrix},
\]
that is, by the Jones vector of the classical field.
Because coherent states remain coherent under linear optics, the output amplitude vector is
$\bs{\alpha}_{\rm out}
=
\alpha\, U c$.
Equivalently, its $m$-th component is
$(\bs\alpha_{\rm out})_m=\alpha\left(U_{m,H}c_H+U_{m,V}c_V\right).$
The output intensity in mode $m$ is then
\[
I_m
=
|(\bs\alpha_{\rm out})_m|^2
=
|\alpha|^2 \left|U_{m,H}c_H+U_{m,V}c_V\right|^2
=
|\alpha|^2 p_m^{(1)}.
\]
Hence, after normalization by the total transmitted intensity,
\begin{equation}
    \tilde p_m
    =
    \frac{I_m}{\sum_{\ell} I_\ell}
    =
    p_m^{(1)},
\end{equation}
so the normalized coherent-state output intensities reproduce exactly the same feature vector as the single-photon output probabilities given in Eq. \ref{eq:single_photon_probability}.

\paragraph{Coherent inputs: two spatial modes.}
For two input spatial modes, the natural classical analogue of the product two-qubit input is a pair of coherent beams, one in each spatial mode, with polarization vectors $c^{(1)}=(c_H^{(1)},c_V^{(1)})^T$ and $c^{(2)}=(c_H^{(2)},c_V^{(2)})^T$. The input coherent amplitudes can be collected into the four-component vector
\[
\boldsymbol{\alpha}_{\rm in}^T
=
(\alpha_1 c_H^{(1)}, \alpha_1 c_V^{(1)}, \alpha_2 c_H^{(2)}, \alpha_2 c_V^{(2)}),
\]
where $\alpha_1,\alpha_2\in\mathbb C$ set the optical powers in the two spatial inputs. After the same linear optical transformation, $\bs\alpha_{\rm out}=U\,\bs\alpha_{\rm in}$, and the output amplitude in mode $m$ is
\begin{equation*}
    (\bs\alpha_{\rm out})_m =
    \sum_{\mu\in\{H,V\}} \left(
        \alpha_1 U_{m,1\mu} c_\mu^{(1)}
        +
        \alpha_2 U_{m,2\mu} c_\mu^{(2)}
    \right),
\end{equation*}
and the corresponding output intensity is
$I_m=|(\bs\alpha_{\rm out})_m|^2$.
Thus in this case each spatial mode carries its own polarization Jones vector, and the reservoir acts linearly on the resulting four-component amplitude vector. However, in the two-qubit case the relation to the quantum output statistics is no longer the same, as was the case for single qubits, and $I_m \not\propto p_{mn}^{(2)}$ in general.

\paragraph{Independent spatial modes.}
Suppose now that the two spatial modes evolve independently. In this case the overall transformation is local, $U = U^{(1)}\otimes U^{(2)}$,
where $U^{(1)}$ acts on the polarization qubit in spatial mode $1$ and $U^{(2)}$ on the one in spatial mode $2$. For a general two-qubit input state
\[
\ket{\Psi_{\rm in}}=\sum_{\mu,\nu\in\{H,V\}} c_{\mu\nu}\,\ket{\mu}_1\ket{\nu}_2,
\]
which may be entangled across the two spatial modes, the output state becomes
\[
    \ket{\Psi_{\rm out}}
    =
    \sum_{m,n}
    \left(
    \sum_{\mu,\nu} c_{\mu\nu}\, U^{(1)}_{m\mu}U^{(2)}_{n\nu}
    \right)
    \ket{m}_1\ket{n}_2.
\]
Hence the joint output probability of obtaining mode $m$ in branch $1$ and mode $n$ in branch $2$ is
\[
    p_{mn}
    =
    \left|
    \sum_{\mu,\nu} c_{\mu\nu}\, U^{(1)}_{m\mu}U^{(2)}_{n\nu}
    \right|^2
    =
    \Tr\!\left[
    \left(\mu_m^{(1)}\otimes \mu_n^{(2)}\right)\rho_{\rm in}
    \right],
\]
with
\[
    \mu_m^{(1)}=(U^{(1)})^\dagger \ket{m}\!\bra{m} U^{(1)},
    \quad
    \mu_n^{(2)}=(U^{(2)})^\dagger \ket{n}\!\bra{n} U^{(2)},
\]
and $\rho_{\rm in}=|\Psi_{\rm in}\rangle\langle\Psi_{\rm in}|$ (or, more generally, an arbitrary two-qubit density matrix). If the input is separable, $c_{\mu\nu}=c_\mu^{(1)}c_\nu^{(2)}$, then the probabilities factorize,
\[
p_{mn}
=
\left| \sum_\mu U^{(1)}_{m\mu} c_\mu^{(1)} \right|^2
\left| \sum_\nu U^{(2)}_{n\nu} c_\nu^{(2)} \right|^2
=
p_m^{(1)} p_n^{(2)}.
\]
By contrast, for entangled inputs the same local evolution generally produces non-factorizing joint probabilities, entirely through the coefficients $c_{\mu\nu}$ of the input state.

In the coherent case, the natural analogue is a product of two coherent beams, one in each spatial branch, with input Jones vectors $c^{(1)}=(c_H^{(1)},c_V^{(1)})^T$ and $c^{(2)}=(c_H^{(2)},c_V^{(2)})^T$:
\[
\boldsymbol{\alpha}_{\rm in}^{(1)}=\alpha_1 c^{(1)},
\qquad
\boldsymbol{\alpha}_{\rm in}^{(2)}=\alpha_2 c^{(2)}.
\]
The two branches evolve independently as
\[
\boldsymbol{\alpha}_{\rm out}^{(1)}=\alpha_1 U^{(1)} c^{(1)},
\qquad
\boldsymbol{\alpha}_{\rm out}^{(2)}=\alpha_2 U^{(2)} c^{(2)},
\]
so that the output intensities are
\[
I_m^{(1)}=|\alpha_1|^2 \left| \sum_\mu U^{(1)}_{m\mu} c_\mu^{(1)} \right|^2,
\,\,
I_n^{(2)}=|\alpha_2|^2 \left| \sum_\nu U^{(2)}_{n\nu} c_\nu^{(2)} \right|^2.
\]
After normalization within each branch, we get
\[
\frac{I_m^{(1)}}{\sum_r I_r^{(1)}}=p_m^{(1)},
\qquad
\frac{I_n^{(2)}}{\sum_s I_s^{(2)}}=p_n^{(2)}.
\]
Thus, when the evolution is factorized, coherent training reproduces exactly the local single-qubit feature maps associated with the two branches. However, a pair of product coherent states encodes only $c^{(1)}$ and $c^{(2)}$ individually, and therefore can only replicate measurement statistics corresponding to separable two-qubit inputs.

\section*{Acknowledgments}

This work is supported by the ERC Advanced Grant QU-BOSS (QUantum advantage via non-linear BOSon Sampling, Grant No. 884676), by the project QU-DICE, Fare Ricerca in Italia, Ministero dell'istruzione e del merito, code: R20TRHTSPA, by the PNRR MUR project PE0000023-NQSTI (Spoke 4), and by the MUR PRIN project QCAPP No. 2022LCEA9Y.
GLM acknowledge funding from the European Union - NextGenerationEU through the Italian Ministry of University and Research under PNRR-M4C2-I1.3 Project PE-00000019 ”HEAL ITALIA” (CUP B73C22001250006). SL, AF, and GMP acknowledge support by MUR under PRIN Project No. 2022FEXLYB. Quantum Reservoir Computing (QuReCo). LI, SL, MP, and GMP acknowledge funding from the “National Centre for HPC, Big Data and Quantum Computing (HPC)” Project CN00000013 HyQELM – SPOKE 10. MP is grateful to the Royal Society Wolfson Fellowship (RSWF/R3/183013), the Department for the Economy of Northern Ireland under the US-Ireland R\&D Partnership Programme, the PNRR PE Italian National Quantum Science and Technology Institute (PE0000023), and the EU Horizon Europe EIC Pathfinder project QuCoM
(GA no. 10032223).




\clearpage
\bibliography{QiSMG16}

\end{document}


\title{\textit{Supplementary Information for}:\\ Efficient classical training of model-free quantum photonic reservoir}

\let\comma,
\author{Rosario Di Bartolo}
\affiliation{Dipartimento di Fisica - Sapienza Università di Roma\comma{} P.le Aldo Moro 5\comma{} I-00185 Roma\comma{} Italy}
\author{Valeria Cimini}
\affiliation{Dipartimento di Fisica - Sapienza Università di Roma\comma{} P.le Aldo Moro 5\comma{} I-00185 Roma\comma{} Italy}
\author{Giorgio Minati}
\affiliation{Dipartimento di Fisica - Sapienza Università di Roma\comma{} P.le Aldo Moro 5\comma{} I-00185 Roma\comma{} Italy}
\author{Danilo Zia}
\affiliation{Dipartimento di Fisica - Sapienza Università di Roma\comma{} P.le Aldo Moro 5\comma{} I-00185 Roma\comma{} Italy}
\author{Luca Innocenti}
\affiliation{Universit\`a degli Studi di Palermo\comma{} Dipartimento di Fisica e Chimica - Emilio Segr\`e\comma{} via Archirafi 36\comma{} I-90123 Palermo\comma{} Italy}
\author{Salvatore Lorenzo}
\affiliation{Universit\`a degli Studi di Palermo\comma{} Dipartimento di Fisica e Chimica - Emilio Segr\`e\comma{} via Archirafi 36\comma{} I-90123 Palermo\comma{} Italy}
\author{Gabriele Lo Monaco}
\affiliation{Universit\`a degli Studi di Palermo\comma{} Dipartimento di Fisica e Chimica - Emilio Segr\`e\comma{} via Archirafi 36\comma{} I-90123 Palermo\comma{} Italy}
\author{Nicol\`{o} Spagnolo}
\affiliation{Dipartimento di Fisica - Sapienza Università di Roma\comma{} P.le Aldo Moro 5\comma{} I-00185 Roma\comma{} Italy}
\author{Taira Giordani}
\affiliation{Dipartimento di Fisica - Sapienza Università di Roma\comma{} P.le Aldo Moro 5\comma{} I-00185 Roma\comma{} Italy}
\author{G. Massimo Palma}
\affiliation{Universit\`a degli Studi di Palermo\comma{} Dipartimento di Fisica e Chimica - Emilio Segr\`e\comma{} via Archirafi 36\comma{} I-90123 Palermo\comma{} Italy}
\author{Mauro Paternostro}
\email{mauro.paternostro@unipa.it}
\affiliation{Universit\`a degli Studi di Palermo\comma{} Dipartimento di Fisica e Chimica - Emilio Segr\`e\comma{} via Archirafi 36\comma{} I-90123 Palermo\comma{} Italy}
\affiliation{Centre for Quantum Materials and Technologies, School of Mathematics and Physics, Queen’s University Belfast, BT7 1NN, United Kingdom}
\author{Alessandro Ferraro}
\affiliation{Dipartimento di Fisica Aldo Pontremoli\comma{} Universit\`a degli Studi di Milano\comma{} I-20133 Milano\comma{} Italy}
\author{Fabio Sciarrino }
\email{fabio.sciarrino@uniroma1.it}
\affiliation{Dipartimento di Fisica - Sapienza Università di Roma\comma{} P.le Aldo Moro 5\comma{} I-00185 Roma\comma{} Italy}

\maketitle

\section{Experimental details}

\subsection{Gradient estimation and optimization protocol}
We adopt a standard (batch) gradient-descent approach in which the parameters $\bm{\nu}=(\theta,\phi)$ are iteratively updated along the negative gradient of the loss function $\mathcal{L}$ \cite{ruder2017overviewgradientdescentoptimization}. In our implementation, each loss evaluation is performed over a finite set of input states, effectively defining a mini-batch. While previous experimental works \cite{suprano2024experimental,zia2025quantum} often relied on large training datasets (150 states) to approximate the full distribution, here we employ a reduced but expressive mini-batch of $15$ randomly sampled states, which provides a favorable compromise between accuracy, generalization capability and experimental overhead.

\textit{Finite-difference gradients.}
Since analytical gradients are not directly accessible in our photonic platform, they are estimated via finite differences. Starting from a given point $(\theta,\phi)$, the loss is evaluated at displaced parameter values, allowing one to reconstruct the local gradient component along each direction. This procedure enables a direct experimental estimation of the gradient using only intensity measurements, without requiring an explicit model of the underlying reservoir dynamics.

\textit{Learning rate.}
The learning rate $\eta$ determines the step size in parameter space and is chosen to ensure stable convergence while minimizing the number of required experimental evaluations. Its value is set empirically by balancing convergence speed against the intrinsic noise of the measurements, and it has been set to 0.8.

\textit{Experimental sensitivity.}
The choice of the finite-difference step $\epsilon$ is dictated by a trade-off between the mechanical resolution of the rotation stages and the sensitivity of the measured signal. The wave plates are mounted on motorized rotation stages with a repeatable incremental motion of $0.03^\circ$ and a bidirectional repeatability of $\pm 0.1^\circ$. Accordingly, the minimum reliable update step is $0.1^\circ$, while the finite-difference increment is set to $\epsilon = 2.87^\circ$, ensuring a measurable variation of the loss while remaining within the stability bounds of the apparatus.

\subsection{Measurement procedure and statistical model}

The overall optimization time is critically influenced by the data acquisition process. For each loss evaluation, both the number of classical samples and the integration time of the power meter are carefully optimized to accurately reconstruct the output probability distributions from intensity measurements, while keeping the acquisition time as short as possible. As described in the main text, the measurement protocol consists of projective measurements in the angular momentum space: the polarization is first selected via a polarizing beam splitter (PBS), then a spatial mode projection is performed using a spatial light modulator and the coupling into a single-mode fiber. Within this framework, the reconstruction of the output probability distribution requires a number of measurement settings equal to the dimension of the explored orbital angular momentum (OAM) space.

In the case of classical light, measurements are performed using a power meter with an averaging parameter set to $n_{\mathrm{samples}} = 10$ and an average acquisition time equal to $\tau = 10$ s. This choice is consistently adopted for both single- and double-quantum-walk configurations. For coherent light inputs, the state is effectively separable, implying that both the evolution and the measurement outcomes factorize across the two paths. In our implementation, the two-step quantum walk (with the q-plate configuration exploring the OAM space $\{m=-2,-1,0,1,2\}$) requires $5$ measurements setting, without increasing the sampling requirements, thus allowing efficient data acquisition.

For quantum measurements, detection is carried out using avalanche photodiodes. 
In this case, each measurement consists of a single repetition with an exposure time of $3\,\mathrm{s}$. 
Using one photon as an heralder and measuring the other leads to sufficiently high counting statistics from the spontaneous parametric down-conversion source, with typical rates of $\sim 10^3\,\mathrm{counts/s}$ at the end of the quantum walk.

In the two-photon case, as in the reconstruction of the entanglement witness, the joint state spans an OAM Hilbert space of dimension $5^2 = 25$, thus requiring a $25$ measurements setting within the same projective scheme. In this regime, the coincidence rate at the end of the double quantum walk decreases to $\sim 30\,\mathrm{counts/s}$, making longer acquisition times necessary. Therefore, each measurement is performed by averaging over multiple repetitions, with an exposure time of $10\,\mathrm{s}$ 
per setting, in order to obtain reliable statistical estimates. The direct consequence of the longer acquisition time for each measurement is that this case is more sensitive to the experimental drifts and imperfections of the apparatus.

Finally, we note that the uncertainty associated with classical intensity measurements depends on photodiode power sensors and is evaluated as the $3\%$ of the measured intensity, valid in the $440\,nm-980\,nm$ wavelength range, divided by $\sqrt{n_{samples} \cdot \tau}$, to account for the number of repeated acquisitions and the integration time of each acquisition. The uncertainties of the training performance are then evaluated by propagating the error under the assumption of Gaussian error propagation.
In contrast, the statistical uncertainty of the experimental results, in particular for the coincidence measurements of photons (as reported in Fig. 3 of the main text and below in Fig. \ref{fig:5}), is quantified by propagating photon-counting shot noise through the full data analysis pipeline using a Monte Carlo resampling procedure. Starting from the experimentally measured photon counts used to reconstruct the output probability distributions, synthetic datasets are generated by independently resampling each count according to a Poisson distribution. This procedure faithfully captures the dominant noise source in the experiment, namely finite counting statistics. For each task, 100 resampled datasets are generated, and the complete performance evaluation is recomputed for each realization, allowing us to extract statistically robust uncertainty estimates.

As a possible route toward improving both the efficiency and stability of the measurement process, one could adopt a detection scheme that minimizes photon losses and reduces its sensitivity to mechanical fluctuations of the setup, such as misalignment and other experimental imperfections. In the current implementation, the use of a PBS intrinsically discards half of the signal. This limitation could be overcome by removing the PBS and employing single-photon cameras capable of simultaneously achieving high speed and high sensitivity imaging of the single-photon state. 
In particular, time-resolving pixelated detectors would enable parallel spatial-mode detection, with each pixel time-stamped relative to the detection of the heralding photon. Such an approach would effectively allow for a single-setting measurement, enabling a more efficient reconstruction of the output probability distributions. This would significantly reduce acquisition times while improving robustness against experimental drifts.

\section{Additional experimental results}

\subsection{Photonic reservoir landscape}

To characterize the dependence of the reconstruction performance on the projection stage of the experimental photonic reservoir, we mapped the reconstruction error over a dense grid of projection settings. 
More specifically, after the reservoir evolution, the output state was projected onto a basis defined by the angles of a half-wave plate (HWP) and a quarter-wave plate (QWP), and the reconstruction task was repeated for each point of the corresponding two-dimensional parameter space. 
This procedure yields an experimental reservoir landscape, where the vertical coordinate represents the average reconstruction error on the training set used and the horizontal coordinates encode the measurement setting used for the projection stage.

\begin{figure*}[ht]
    \centering \includegraphics[width=0.99\linewidth]{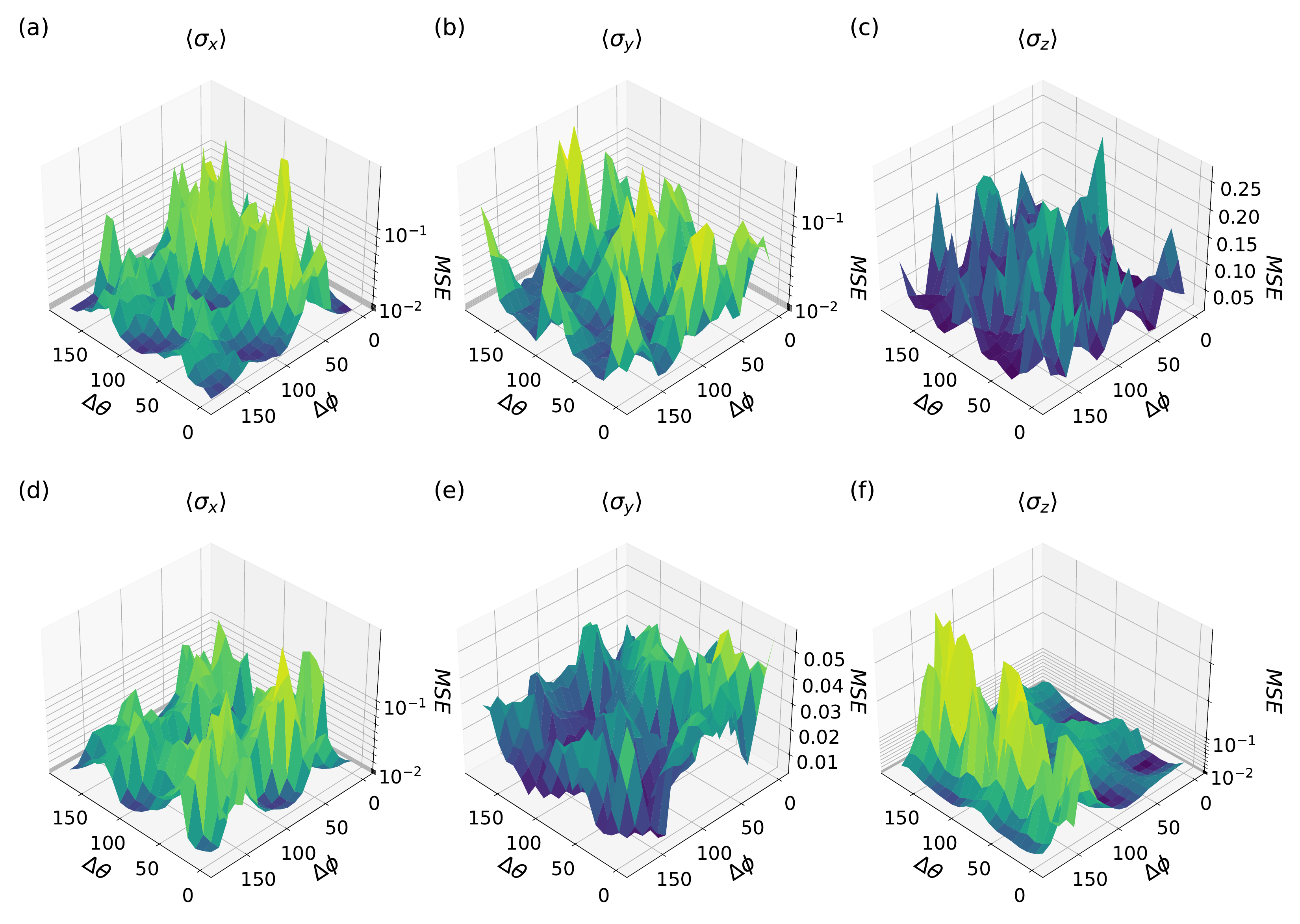}
    \caption{\textbf{Experimental performance landscapes for Pauli observable reconstruction with a quantum reservoir implemented via a single quantum walk and coherent classical light.}
    Each landscape is obtained by scanning the projection measurement stage over a $20 \times 20$ grid of half-wave plate and quarter-wave plate angles ($\theta$ and $\phi$), both varied from $0^\circ$ to $180^\circ$ with a step equal to $\Delta=9.47^\circ$, where $0^\circ$ corresponds to the theoretical model parameter and all other values represent a shift from this, as shown on the horizontal axes. 
    The vertical axis reports the mean squared error associated with the reconstruction of the target observable. 
    The dataset consists of 15 randomly prepared polarization states of coherent light, each initialized in the horizontal polarization state and then transformed into a random input state before undergoing the reservoir evolution. 
    For every point of the grid, the same evolved output states are measured in a different projection basis, thereby probing how the reconstruction performance depends on the reservoir parameters, and in particular on the projection setting. 
    Panels (a)-(c) and (d)-(f) report two measurements performed on the same dataset under different alignment conditions, showing how the resulting performance landscapes can change even for the same reservoir architecture and input dataset. 
    The reconstructed observables are $\langle \sigma_x \rangle$ in panels (a) and (d), $\langle \sigma_y \rangle$ in panels (b) and (e), and $\langle \sigma_z \rangle$ in panels (c) and (f). Additionally, focusing on $\langle \sigma_y \rangle$ and panel (b), the MSE evaluated at the theoretical point $(\theta=0^\circ,\phi=0^\circ)$ is $(5.201 \pm 0.069)\times 10^{-2}$, whereas the minimum identified by grid search occurs at $(\theta=28.42^\circ,\, \phi=0.0^\circ)$, yielding an MSE of $(3.588 \pm 0.019)\times 10^{-3}$. Similarly, for panel (e), the MSE at the theoretical point $(\theta=0^\circ,\phi=0^\circ)$ is $(5.588 \pm 0.043)\times 10^{-2}$, while the optimal point found on the grid is $(\theta=9.47^\circ,\, \phi=132.63^\circ)$, corresponding to an MSE of $(2.4258 \pm 0.0073)\times 10^{-3}$.
    \label{fig:landscape_pauli}}
\end{figure*}

The first set of landscapes, reported in Fig.~\ref{fig:landscape_pauli}, concerns the reconstruction of single-qubit Pauli observables using a reservoir implemented through a single quantum walk. In this case, the training dataset consists of 15 randomly prepared polarization states of coherent light, a number chosen to ensure a representative characterization of the experimental properties relevant to the learning task \cite{suprano2024experimental}. Each state is initialized in horizontal polarization, transformed into a random input state, and then injected into the reservoir.
For each point of a $20 \times 20$ grid spanning the HWP and QWP angles ($\theta$ and $\phi$) from $0^\circ$ to $180^\circ$ with a step equal to $\Delta=9.47^\circ$, the same evolved output states are measured in a different projection basis. Here, the point $(\theta=0^\circ,\phi=0^\circ)$ corresponds to the measurement projection obtained by leveraging on the theoretical description of the underlying evolution, while all other points represent angular shifts with respect to them. For each point of the grid, the reconstruction performance is quantified through the mean squared error (MSE) evaluated on the training set. 
The resulting landscapes report the MSE associated with the reconstruction of $\langle \sigma_x \rangle$, $\langle \sigma_y \rangle$, and $\langle \sigma_z \rangle$. These measurements explicitly show that the reconstruction performance is not uniform across the readout parameter space. Rather, the choice of measurement basis affects the quality of the learned mapping between reservoir outputs and target observables.
Indeed, considering $\langle \sigma_y \rangle$, the MSE at the theoretical point $(\theta=0^\circ,\phi=0^\circ)$ is $(5.201 \pm 0.069)\times 10^{-2}$, whereas the minimum identified by grid search occurs at $(\theta=28.42^\circ,\, \phi=0.0^\circ)$, and yields an MSE of $(3.588 \pm 0.019)\times 10^{-3}$ (see Fig. \ref{fig:landscape_pauli}(b)).\\
To further probe the robustness of this dependence, we repeated the acquisition on the same dataset under a different alignment condition of the setup. The corresponding landscapes, shown in panels (d)-(f) of Fig. \ref{fig:landscape_pauli}, demonstrate that the detailed structure of the performance landscape may vary significantly even when the reservoir architecture and the input dataset are kept fixed. This indicates that the experimentally accessible reservoir response is shaped not only by the underlying evolution, but also by the effective optical configuration of the setup. In this alignment condition, while the MSE at the theoretical point $(\theta=0^\circ,\phi=0^\circ)$ is $(5.588 \pm 0.043)\times 10^{-2}$, while the optimal point found on the grid is $(\theta=9.47^\circ,\, \phi=132.63^\circ)$, reaching an MSE of $(2.4258 \pm 0.0073)\times 10^{-3}$ (see Fig. \ref{fig:landscape_pauli}(e)). 

\begin{figure*}[ht]
    \centering \includegraphics[width=0.99\linewidth]{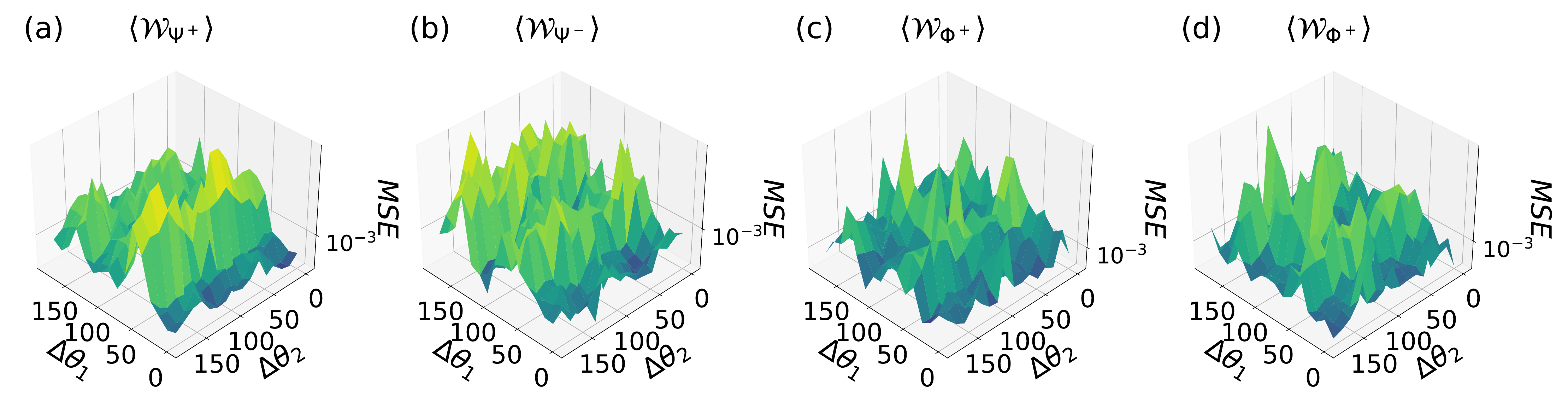}
    \caption{\textbf{Experimental performance landscapes for entanglement witnessing using a double quantum walk and coherent classical light.}
    The landscapes are generated by varying the projection measurement stage over a $16 \times 16$ grid of both half-wave plate settings ($\theta_{1}$ and $\theta_{2}$) of the two lines of the double quantum walk, which define the horizontal axes of the plots. The grid spans from $0^\circ$ to $180^\circ$ with a step equal to $\Delta=12^\circ$, where $0^\circ$ corresponds to the theoretical model parameter and all other values represent a shift from this.
    For each grid point, the mean squared error in the reconstruction of the target witness expectation value is reported along the vertical axis. 
    The dataset is composed of 35 randomly prepared polarization states of coherent light, all initialized in the horizontal polarization state and then transformed into random input states before being injected into the reservoir. 
    After the double quantum walk evolution, the same output states are analyzed under different projection settings, thereby probing the dependence of the witness reconstruction performance on the readout basis. 
    Panels (a)-(d) correspond to the reconstruction of the entanglement witnesses $\langle \mathcal{W}_{\Psi^+} \rangle$, $\langle \mathcal{W}_{\Psi^-} \rangle$, $\langle \mathcal{W}_{\Phi^+} \rangle$, and $\langle \mathcal{W}_{\Phi^-} \rangle$, respectively. Additionally, considering $\langle \mathcal{W}_{\Psi^+} \rangle$, in the theoretical point $(\theta_{1}=0^\circ, \theta_{2}=0^\circ)$ the MSE is equal to $(5.16 \pm 0.37) \times 10^{-4}$, while the best point found by minimizing over the grid is $(\theta_{1}=12.0^\circ, \theta_{2}=108.0^\circ)$ and yields an MSE of $(1.53 \pm 0.30) \times 10^{-4}$. 
    \label{fig:landscape_witness}}
\end{figure*}

The study of the reservoir landscape has been done also for the double quantum walk setup and it is shown in Fig. \ref{fig:landscape_witness}. This configuration enables the exploration of two-photon input states, in which each quantum walk line is associated with the evolution of one photon. The dynamical structure of this setup naturally leads to learning tasks aimed at reconstructing properties of bipartite quantum systems and, more importantly, observables that capture the effect of quantum correlations, such as entanglement witnesses.
In this case, we emphasize the strategy of classical-to-quantum transfer learning by using coherent classical light for the evaluation of entanglement witnesses, while, as discussed in the main text, the final test is performed on entangled states of light. Here, to effectively learn the experimental properties of the system, we need a larger training dataset consisting of 35 randomly prepared polarization states of coherent light. Each state is initialized in horizontal polarization and then its state is modified through 4 wave plates, two in each quantum walk, 
before being injected into the reservoir.
Since the number of reservoir parameters is now doubled, in order to avoid an excessive optimization overhead, we restrict the landscape analysis to one projection angle per quantum walk in the measurement stage. Specifically, to optimize both quantum walk lines, we keep the projection QWPs of the two lines fixed at the theoretical values predicted by the physics-informed model,i.e. ($\phi_1=0^\circ,\phi_2=0^\circ$), and vary only the HWPs angles. To explore the landscape, these angles are scanned over a $16 \times 16$ grid of HWPs settings ($\theta_{1}$ and $\theta_{2}$), spanning from $0^\circ$ to $180^\circ$, with a step equal to $\Delta=12^\circ$. As before, the point $(\theta_{1}=0^\circ,\theta_{2}=0^\circ)$ corresponds to the theoretical model parameters, while all other points represent angular shift relative to them.
For each point of the grid, the reconstruction performance is quantified through the MSE, evaluated on the training set.
For each grid point, the reconstruction error is evaluated for the four target witness observables 
$\langle \mathcal{W}_{\Psi^+} \rangle$, $\langle \mathcal{W}_{\Psi^-} \rangle$, 
$\langle \mathcal{W}_{\Phi^+} \rangle$, and $\langle \mathcal{W}_{\Phi^-} \rangle$, 
where $\{ \Psi^+, \Psi^-, \Phi^+, \Phi^- \}$ denotes the Bell basis, 
with witnesses defined as in the main text.
Also in this case, the resulting landscapes reveal a pronounced dependence of the reconstruction MSE on the choice of projection basis at the output. 
Indeed, while at the theoretical point $(\theta_{1}=0^\circ, \theta_{2}=0^\circ)$ the MSE is equal to $(5.16 \pm 0.37) \times 10^{-4}$, while the best point found by minimizing over the grid is $(\theta_{1}=12.0^\circ, \theta_{2}=108.0^\circ)$, yielding an MSE of $(1.53 \pm 0.30) \times 10^{-4}$ (see Fig. \ref{fig:landscape_witness}(a)). 
This highlights that, even when considering the double quantum walk architecture, the HWPs of the projection stage act as an effective control layer of the reservoir, whose tuning can significantly influence the quality of the reconstructed observables. This also suggests that including QWPs as additional control parameters, which was not done here for representational purposes, in order to achieve full control over the projection measurement, could further improve the search for better minima.

More broadly, the experimentally measured landscapes provide a direct picture of how the information processed by the reservoir is distributed across the accessible parameter space, thereby identifying regions associated with either enhanced or degraded learning performance. At the same time, the marked dependence of these landscapes on the specific alignment condition shows that the experimentally relevant minima are not fixed once and for all, but rather emerge from the actual physical configuration of the setup. As a consequence, the optimal operating points cannot, in general, be pre-calibrated and must instead be identified directly under the actual experimental conditions of each realization. In this respect, the presence of well-defined minima and maxima in the measured performance landscapes is particularly relevant, as it shows that the reservoir parameter space retains a sufficiently structured geometry to be efficiently explored through a gradient-descent-based strategy. This, in turn, enables the online optimization of a model-free quantum photonic reservoir with only a limited number of experimental iterations.

%
%
%
%
%
%
%
%

%
%
%

\subsection{Validation on non-optimized observables}

\begin{figure*}[ht]
    \centering \includegraphics[width=0.99\linewidth]{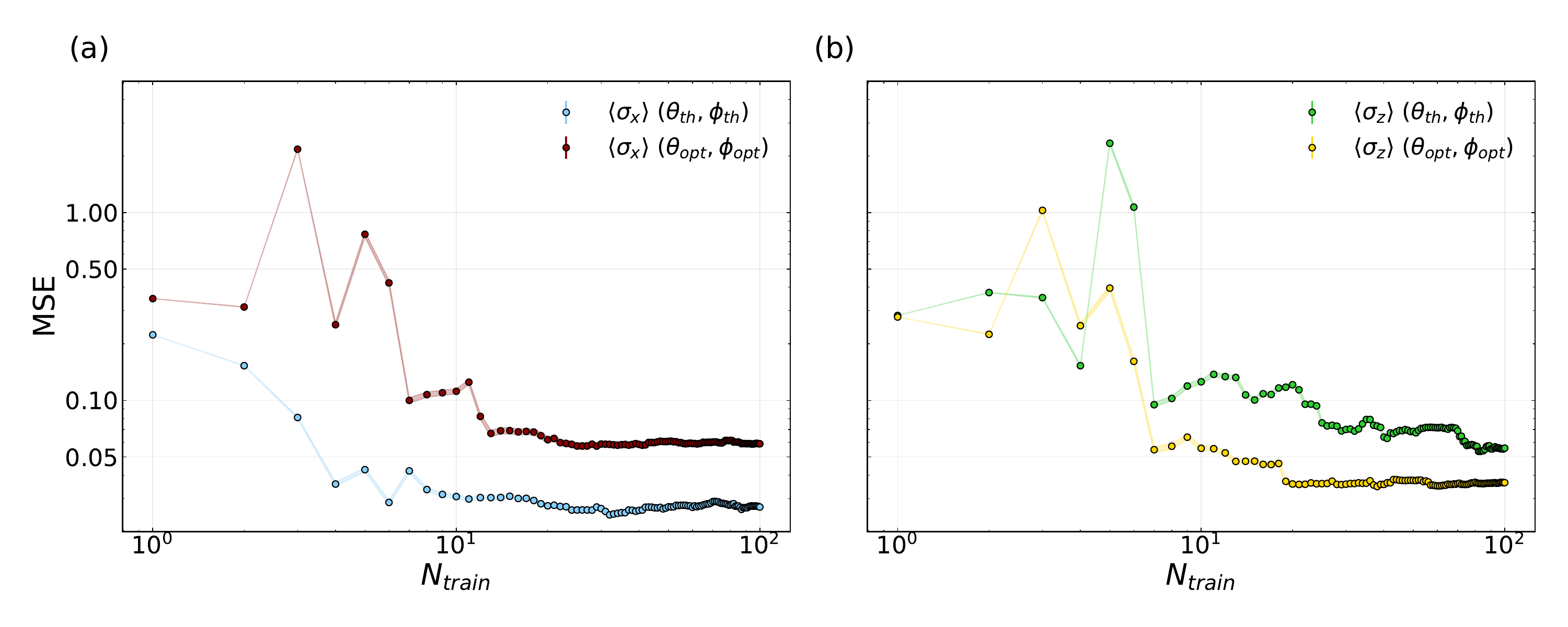}
    \caption{
    \textbf{Experimental learning curves for $\langle \sigma_x \rangle$ and $\langle \sigma_z \rangle$ reconstruction with theoretical and experimental optimized projection angles of the reservoir.} The (a) $\langle \sigma_x \rangle$ and (b) $\langle \sigma_z \rangle$ reconstruction is performed in two reservoir configurations: using the angles optimized according the theoretical model of the photonic platform ($\theta_{th}$, $\phi_{th}$), shown in blue, and according the experimental gradient descent ($\theta_{opt}$, $\phi_{opt}$), shown in red. The training dataset consists of 100 randomly prepared coherent light input states, while the test dataset consists of 100 single-photon input states. All states are the same for the two configurations. The error over the curves is evaluated via 100 realizations of a Monte Carlo Poisson resampling of the experimental photon-count distributions, is reported in the plot as shaded areas and represent one standard deviation of the resulting performance distribution.
    \label{fig:5}}
\end{figure*}

To further assess the robustness of the gradient-descent optimization, we investigate whether measurement settings optimized for a specific observable, that is $\langle \sigma_y \rangle$, remain effective when applied to the reconstruction of different, non-optimized observables, such as $\langle \sigma_x \rangle$ and $\langle \sigma_z \rangle$. This analysis provides a direct test of the generalization capability of the learned measurement configuration and probes whether the optimization captures observable-independent features of the reservoir dynamics.
Using the same training procedure described above, we evaluate the reconstruction performance for the expectation values of the Pauli operators $\langle \sigma_x \rangle$ and $\langle \sigma_z \rangle$. The models are trained exclusively on classical coherent states, while performance is assessed on a test set of $100$ single-photon polarization states. Fig. \ref{fig:5} reports the learning curves as a function of the number of classical training examples $N_{train}$, comparing projection angles derived from the theoretical model $(\theta_{th}, \phi_{th})$ with those obtained via experimental gradient-descent optimization $(\theta_{opt}, \phi_{opt})$. In particular, while in the reconstruction of $\langle \sigma_x \rangle$ the model-informed optimized reservoir achieves better performance, the gradient-descent–optimized configuration systematically attains lower mean squared error values and exhibits a more stable convergence behavior for $\langle \sigma_z \rangle$, showing how the performance advantage of the optimized configuration persists even though the considered observables were not directly targeted during the optimization procedure. This result indicates that the gradient descent algorithm does not simply tailor the measurement settings to a single observable, but rather identifies projections that enhance the overall informativeness of the reservoir response.
For both observables, the reconstruction error decreases as the size of the training set increases, confirming that the reservoir representation enables progressive improvement of the learned readout.
These findings demonstrate that the experimentally optimized measurement configuration generalizes beyond the selected optimization target, enabling accurate reconstruction of multiple Pauli operators without additional tuning. 
Moreover, they provide strong evidence that the learning procedure effectively adapts the measurement basis to the true experimental response of the photonic reservoir, improving robustness against noise and model imperfections while preserving transferability across observables.

\subsection{Robustness of the protocol}

\begin{figure*}[ht]
    \centering
    \includegraphics[width=0.495\linewidth]{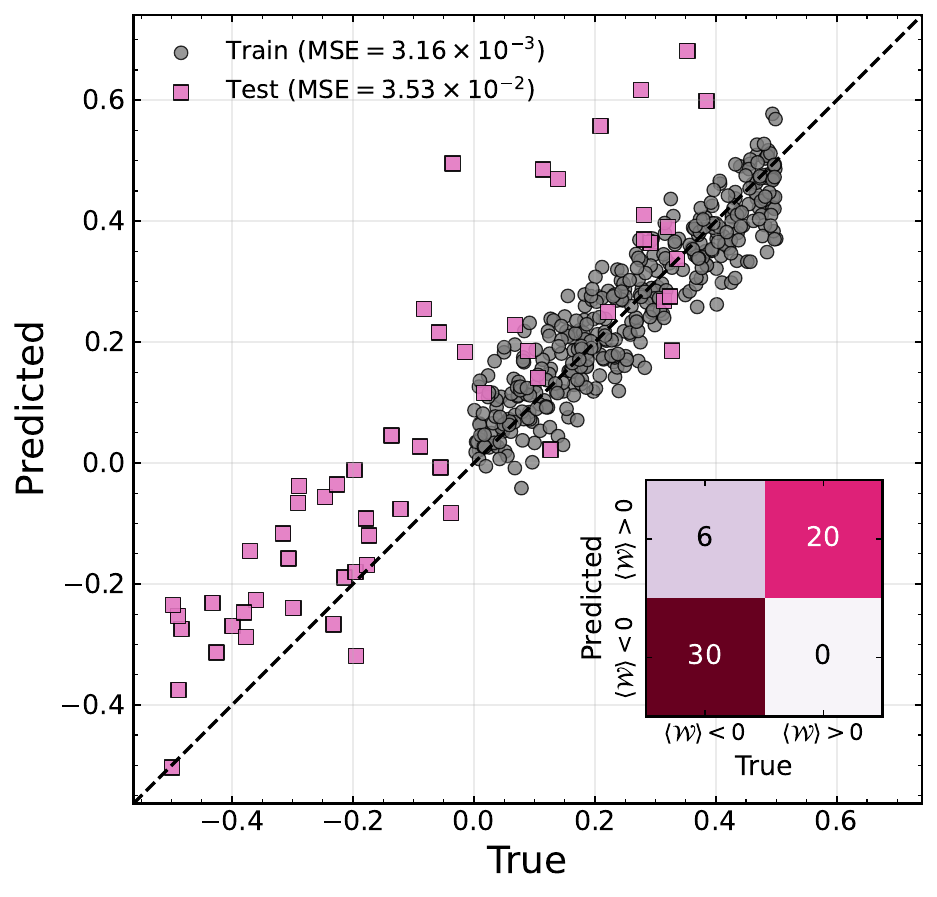}
    \caption{
    \textbf{Experimental analysis of the robustness of the transfer learning protocol.}
    Performing the entanglement witnessing task, we study the robustness of the protocol by repeating the protocol with different experimental conditions (different alignment in different days) on a different dataset and visualize the results by plotting the predicted versus true values. The training set consists of $400$ classical polarization states (grey), while the test set is composed by $56$ entangled states (pink). The mean squared errors for both sets are reported in the legend, while the inset shows the confusion matrix obtained for the entangled test states.
    \label{fig:witness_robustness_SI}}
\end{figure*}

The robustness of the transfer learning protocol is investigated by applying the entanglement witnessing task to another measurement, consisting of different classical training states and different quantum test states. The dataset acquisition is the same as described in the main text, the training set consists of classical coherent states (400 states), while the test set consists of bipartite entangled states (56 states).
The results, shown in Fig. \ref{fig:witness_robustness_SI}, demonstrate that the model achieves performance comparable to the one reported in Fig. 5 of the main text, both in terms of MSE and classification accuracy. Importantly, this confirms that the transfer learning protocol remains valid under an alternative datasets, under varying experimental conditions, without loss of accuracy across different experimental conditions.

\clearpage
\bibliography{QiSMG16}